\documentclass[trackchanges]{aastex631}

\usepackage{multirow}
\usepackage{tabularx}
\usepackage{amsmath,amstext}
\usepackage{bm}
\usepackage{mathtools}
\newcolumntype{Y}{>{\centering\arraybackslash}X}

\shorttitle{MHD model of cloudlet encounter event}
\shortauthors{Unno, Hanawa \& Takasao}
\graphicspath{{./}{figures/}}
\begin{document}

\title{Magnetohydrodynamic Model of Late Accretion onto a Protoplanetary Disk: \\Cloudlet Encounter Event}

\correspondingauthor{Masaki Unno}
\email{unno@astro-osaka.jp}

\author{Masaki Unno}
\affiliation{Department of Earth and Space Science, Graduate School of Science, Osaka University, Toyonaka, Osaka 560-0043, Japan}

\author[0000-0002-7538-581X]{Tomoyuki Hanawa}
\affiliation{Center for Frontier Science, Chiba University, 1-33 Yayoi-cho, Inage-ku, Chiba 263-8522, Japan}

\author[0000-0003-3882-3945]{Shinsuke Takasao}
\affiliation{Department of Earth and Space Science, Graduate School of Science, Osaka University, Toyonaka, Osaka 560-0043, Japan}

\begin{abstract}
Recent observations suggest late accretion, which is generally nonaxisymmetric, onto protoplanetary disks. We investigated nonaxisymmetric late accretion considering the effects of magnetic fields. Our model assumes a cloudlet encounter event at a few hundred au scale, where a magnetized gas clump (cloudlet) encounters a protoplanetary disk. We studied how the cloudlet size and the magnetic field strength affect the rotational velocity profile in the disk after the cloudlet encounter. The results show that a magnetic field can either decelerate or accelerate the rotational motion of the cloudlet material, primarily depending on the relative size of the cloudlet to the disk thickness. When the cloudlet size is comparable to or smaller than the disk thickness, magnetic fields only decelerate the rotation of the colliding cloudlet material. However, if the cloudlet size is larger than the disk thickness, the colliding cloudlet material can be super-Keplerian as a result of magnetic acceleration. We found that the vertical velocity shear of the cloudlet produces a magnetic tension force that increases the rotational velocity. The acceleration mechanism operates when the initial plasma $\beta$ is $ \beta \la 2\times 10^1 $. Our study shows that magnetic fields modify the properties of spirals formed by tidal effects. These findings may be important for interpreting observations of late accretion.

\end{abstract}

\def\Rnum#1{\resizebox{0.5em}{\height}{\uppercase\expandafter{\romannumeral #1}}}

\section{Introduction} \label{sec:intro}
Protoplanetary disks are natural byproducts of the formation of stars through the gravitational collapse of rotating gas clouds. Direct three-dimensional (3D) magnetohydrodynamic (MHD) simulations of gravitational collapse of molecular cloud cores revealed several fundamental processes of formation and growth of disks \citep[e.g.][]{machida2007ApJ,hennebelle2009,Tomida2013ApJ} and driving jets and outflows \citep[e.g.][]{Banerjee2006,Machida2008ApJ}. Previous studies have found that magnetic fields play essential roles in removing angular momentum from accreting gas at different evolutionary stages and radii \citep[e.g.][]{Mouschovias1976ApJ,Tomisaka2002ApJ,Hennebelle2008A&A}. The deceleration of rotating gas by magnetic tension is called magnetic braking.

Theoretical studies of star and disk formation have been progressing along with advances in numerical simulations. Spherically symmetric models were used to study the gravitational collapse of molecular clouds \citep[e.g.][]{Larson1969MNRAS,Masunaga2000ApJ}. The formation of protoplanetary disks has been investigated using a gravitationally unstable, isolated rotating molecular cloud with a spherically symmetric density structure as the initial condition \citep[e.g., see reviews by][]{Inutsuka2012PTEP,Li2014prpl}.
With this type of setup, accretion onto the disks occurs nearly in an axisymmetric manner and becomes monotonically weaker with time.
Numerical simulations that relax the assumption of axisymmetry have been performed to study the effects of the misalignment between the rotational axis and the background magnetic fields in an axisymmetric density structure \citep[e.g.][]{matsumoto2004ApJ,Joos2012A&A,Tsukamoto2015MNRAS}. Recently, more complicated star and disk formation processes have been investigated that consider the effects of turbulence on the core scale or larger \citep[e.g.][]{Joos2013A&A,Lee2016A&A,kuffmeier2017,Lam2019MNRAS}.

Heterogeneous late infall will commonly occur because star-forming regions are intrinsically inhomogeneous. Star-forming molecular clouds must be turbulent \citep[see the review by][]{hennebelle2012}. \citet{kuffmeier2017} performed 3D MHD simulations of star formation starting from the giant molecular cloud (GMC)-scale and showed that the accretion rate onto the disk significantly varies with time, depending on the condition of the surrounding environment. 
They also found accretion of gas clumps or ``cloudlets" onto a disk \citep{kuffmeier2018}.
Star-disk systems formed in turbulent regions will naturally have a finite relative velocity to the ambient gas, which is another important factor to determine the accretion rate. The accretion rate becomes higher when star-disk systems enter higher density regions or with an increase in the infall rate. The importance has been extensively studied based on Bondi-Hoyle-Lyttleton accretion \citep{padoan2005,throop2008,moeckel2009,scicluna2014,wijnen2016}.
From the observational census toward NGC 3603, \citet{beccari2010} argued the necessity of late infall in old ($>$10 Myr) pre-main sequence stars with disks.

Some observations support the idea of the heterogeneous accretion onto disks.
Tail structures connecting to disks with a size of 100-1000 au are found in several systems such as AB Aur \citep{nakajima1995,grady1999}, HD 100546 \citep{ardila2007}, Z CMa \citep{nakajima1995,liu2016}, SU Aur \citep{akiyama2019,ginski2021}, and RU Lup \citep{Huang2020}. These asymmetric structures may be the result of late infall from the remnants of envelopes or giant molecular clouds.
The sulfur monoxide (SO) emissions provide evidence that asymmetrically infalling gas forms accretion shocks around disks \citep{sakai2016,garufi2021}. Those observations suggest that late infall onto disks will be common and that disks will be subject to asymmetric accretion. Moreover, connection to the origin of the misalignment of inner and outer disk regions has been discussed observationally \citep{ginski2021}, which could be consistent with hydrodynamic models \citep{2011MNRAS.417.1817T,2021A&A...656A.161K}.

Motivated by observations and simulations, 3D hydrodynamic simulations of late infall onto protoplanetary disks have been performed to study the detailed process of late encounter events. \citet{dullemond2019} investigated the late encounter between a star and a cloudlet on a scale of several thousand au and showed the encounter can lead to the formation of arc or tail-like structures. \citet{kuffmeier2020} also performed a set of hydrodynamic simulations of the cloudlet encounter on a similar scale and studied the formation of second-generation disks. These 1000 au scale simulations highlight how the stellar gravity captures GMC scale gas. However, detailed interaction between infalling gas and pre-existing protoplanetary disks on a 10-100 au scale remains unclear. Magnetic fields play important roles at this scale because the cloudlets strongly bend and amplify magnetic fields during the encounter and increase the importance of magnetic tension.
Because the encounter process at this scale determines the mass and angular momentum supply to the disks, detailed investigations based on MHD models are necessary.

By performing a set of 3D MHD simulations, we investigate the magnetic effects during the cloudlet encounter event on several 100 au scale. The rest of this paper is structured as follows: Section~\ref{sec:model} describes our model setup. The disk and cloudlet structures are explained. Section~\ref{Result} presents numerical results. The role of magnetic tension is clarified by varying the initial size of a cloudlet. We explain that magnetic tension can not only decelerate but also accelerate the rotation motion of infalling gas depending on the size of the encountering cloudlet.
In addition, a brief comparison to a hydrodynamic model is presented. Section~\ref{sec:discussion} will briefly discuss the magnetic field strength of the cloudlet required for magnetic acceleration and the impact of the non-ideal MHD effect. Section \ref{sec:summary} summarizes our results.

\section{Model}\label{sec:model}
\subsection{Basic Equations}
We performed 3D MHD simulations to examine the nonaxisymmetric accretion process of a magnetized gas clump (cloudlet) onto a protoplanetary disk (as shown in Section~\ref{sec:model_setup}).
It is assumed that the cloudlet and disk consist of cold molecular gas. The cold cloudlet and disk are surrounded by a warm neutral atomic gas.  
We solve the following ideal MHD equations, 
\def\vec#1{\mbox{\boldmath $#1$}}
\begin{align}
\frac{\partial\rho}{\partial t}+\vec{\nabla}\cdot(\rho\vec{v}) &= 0 \label{mass} ,\\
\frac{\partial\rho\vec{v}}{\partial t} +  \vec{\nabla}\cdot\left[\rho\vec{v}\vec{v} + p_t\vec{I} - \frac{\vec{B}\vec{B}}{4\pi}\right] &= -\rho\vec{\nabla}\Phi , \label{momentum}\\
\frac{\partial\vec{B}}{\partial t} + \vec{\nabla}\cdot\left(\vec{v}\vec{B} - \vec{B}\vec{v}\right) &= \bf{0} ,\label{induction}\\
\frac{\partial e}{\partial t} + \vec{\nabla}\cdot\left[\left(e + p_t\right)\vec{v} - \frac{\vec{B}}{4\pi}\left(\vec{v}\cdot\vec{B}\right)\right] &=-\rho \bm{v}\cdot\bm{\nabla}\Phi , \label{energy eq}\\
e = \frac{p}{\gamma - 1} + \frac{1}{2}\rho\left|\vec{v}\right|^2 + \frac{\left|\vec{B}\right|^2}{8\pi} ,\label{energy}\\
p_t = p + \frac{\left|\vec{B}\right|^2}{8\pi} , \label{pressure}
% p = R\rho T ,\label{eos}\\
% R = \frac{k_B}{\mu m_{\rm H}} ,\label{gas constant}
\end{align}
where $\rho,~p,~\bm{v},~\bm{B}$ and $\Phi$ denote the mass density, the thermal pressure, the velocity vector, the magnetic field vector, and the gravitational potential by the protostar, respectively.
The equation of state is 
\begin{eqnarray}
p & = & \frac{\rho k _{\rm B}  T }{\mu m _{\rm H}} ,\label{eos}
\end{eqnarray}
where $ T $, $ k _{\rm B} $, and $ m _{\rm H} $ denote the temperature, the Boltzmann constant, and the mass of a hydrogen
atom, respectively.
We assume that the gas is nearly isothermal, considering that the thermal relaxation timescale is significantly shorter than the dynamical timescale \citep[e.g.][]{Aota_2015}. For this reason, we adopt an ideal gas that has a specific heat ratio, $\gamma = 1.05$. We ignore explicit heating and cooling in this study. Non-ideal MHD effects such as ambipolar diffusion are also ignored. We will discuss the validity of this assumption in section \ref{Non-ideal}.

The source of gravity in our model is only the (spatially unresolved) central protostar with a mass of $M=0.5 M_\odot$. We ignore the self-gravity of the gas disk and the cloudlet. The gravitational potential of the protostar is softened artificially within a radius of $a_{\rm s}$ from the protostar. The functional form is expressed as

\begin{eqnarray}
\Phi(r,z) = 
\begin{dcases}
-\frac{GM}{a_{\rm s}}\left[\frac{3}{2} - \frac{1}{2}\left(\frac{r^2+z^2}{a_{\rm s}^2}\right)\right] &~{\rm if~} \sqrt{r^2 + z^2} < a_{\rm s} \\
-\frac{GM}{a_{\rm s}}\left(\frac{a_{\rm s}^2}{r^2+z^2}\right)^{1/2}&~\rm otherwise
\end{dcases}
\label{g},
\end{eqnarray}
in the cylindrical coordinates, $ (r, \varphi, z) $, with the origin at the central protostar. 
A smaller softening length $a_{\rm s}$ requires a larger computational resources. Due to the limitation of our resources, we take $a_{\rm s} = 100~\rm au$.

The softening of the gravitational potential introduces some artificial structures including a ring-like structure around $r=a_{\rm s}$ in the density map. However, we argue in Appendix \ref{appendix: softening} that the cloud-disk interaction is not significantly affected by the softening.

\subsection{Model Setup}\label{sec:model_setup}
\begin{figure}
\plotone{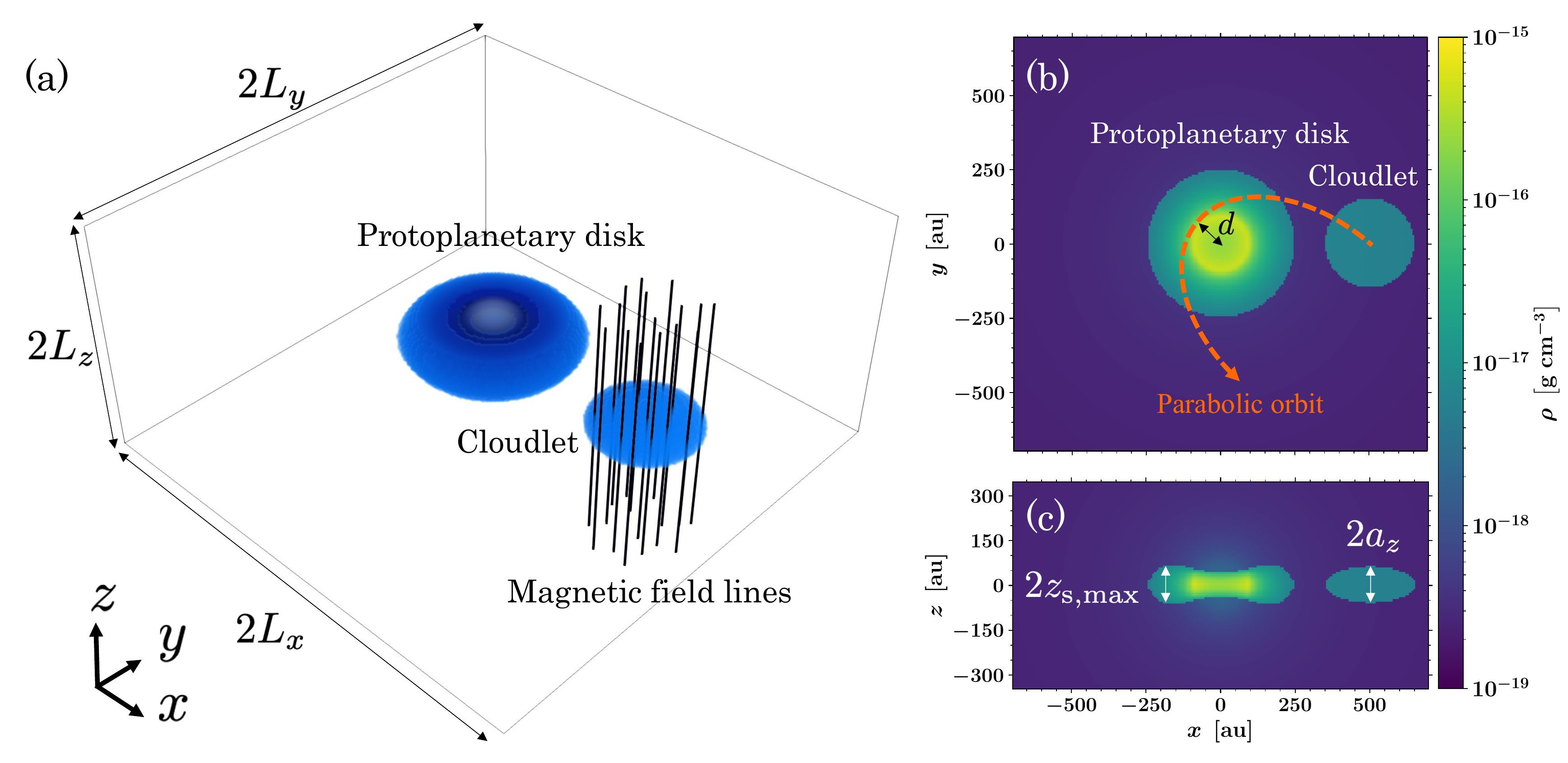}
\caption{Our model setup. Panel (a) shows the three-dimensional structures of the protoplanetary disk, cloudlet, and magnetic field lines. The magnetic field is imposed only on the cloudlet. Panel (b) shows the midplane ($z = 3.5~{\rm au}$) slice. The Orange dashed arrow indicates the parabolic orbit of the cloudlet. Panel (c) displays the cutout at $y = 3.5~{\rm au}$. Color in panels (b) and (c) denotes the density. 
\label{schematic view}}
\end{figure}

Figure~\ref{schematic view} displays the top and side views of the initial setting. The protoplanetary disk and cloudlet are surrounded by a warm neutral medium.

\subsubsection{Warm neutral medium}\label{wnm}

The warm neutral medium is assumed to be in hydrostatic equilibrium under the gravitational potential given by Equation (\ref{g}).
The pressure and density distributions outside the disk and the cloudlet are given as
\begin{eqnarray}
p(r,z)&=&p_0 \exp\left[ -\frac{\mu _0 m _{\rm H} \Phi(r,z)}{k _{\rm B} T_0}\right] \label{pn},\\
\rho(r,z)&=&\rho_0\exp\left[ -\frac{\mu _0 m _{\rm H} \Phi(r,z)}{k _{\rm B} T_0}\right]\label{rhon}, 
%p _0 & = & \frac{k _{\rm B} T _0}{\mu _0 m _{\rm H}} \rho _0,
\end{eqnarray}
where $p_0$ and $\rho_0$ denote the pressure and density at a large distance from the star, respectively. We take $\rho_0 = 2.0\times10^{-19}~{\rm g~cm^{-3}}$. 
The temperature and mean molecular weight of the warm neutral medium are assumed to be $T _{\rm 0} =410~ \rm K $ and $ \mu _{\rm 0} = 1.27 $, respectively. Substituting these values into Equation (\ref{eos}), we obtain $p_0 = 5.3\times 10^{-9} \rm {erg~cm^{-3}}$.

\subsubsection{Protoplanetary disk}\label{PPD}
The disk is set in the region of $r \le R_{\rm d}$ and $|z| \le z_{\rm s}\left(r\right)$, where $R_{\rm d}$ and $z_{\rm s}(r)$ denote the disk outer radius and the disk upper boundary, respectively. We also set the length unit $r_0$ as $100~\rm au$. The upper boundary is defined as

\begin{eqnarray}
z_{{\rm s}} (r) =
\frac{2}{5} \sqrt{r_0 ^2 + r ^2} \,{{\rm tanh}} \displaystyle \left[\frac{3}{2}\left( \frac{R _{\rm d} - r}{r_0}\right)\right] .\label{zs}
\end{eqnarray}
The disk outer radius is set to $ R _{d} = (5/2) r_0 = 250 ~{\rm au}$. The disk is mildly flared and has the maximum height, $ z _{\rm s, max} = 66$ au at $ r= 163 $ au.

The disk has a uniform temperature of $ T _{\rm d} = 49~\mbox{K} $, if the mean molecular weight is $ \mu _{\rm d} = 2.3 $. To ensure the pressure balance with the warm neutral medium, we impose the following boundary condition:
\begin{eqnarray}
p [r, z _{\rm s} (r)] & = & p_0{{\rm exp}}\left\{-\frac{\mu _0 m _{\rm H} \Phi\left[r,z_{{\rm s}}(r)\right]}{k _{\rm B} T_0}\right\}
= p_{{\rm s}} (r) . \label{ps}
\end{eqnarray}
Combining the hydrostatic equations and Equation (\ref{ps}), we obtain 
\begin{eqnarray}
p(r,z)=p_{{\rm s}} (r) \exp \left\{\frac{\mu _{\rm d} m _{\rm H} [\Phi[r,z_{{\rm s}}(r)]-\Phi(r,z)]}{k  _{\rm B} T_{\rm d}}\right\} .\label{pd}
\end{eqnarray}
Substituting Equation (\ref{pd}) into the radial component of the hydrostatic equation, we obtain 
\begin{eqnarray}
v_\varphi[r,z_{\rm s}(r)]^2=r\left(1-\frac{\mu_{{\rm 0}} T_{\rm d}}{\mu _{\rm d} T_0}\right)\left\{ \frac{\partial\Phi\left[r,z_{\rm s}(r) \right]}{\partial r} +\frac{dz_{\rm s}(r)}{dr}\frac{\partial\Phi\left[r,z_{\rm s}(r)\right]}{\partial z_{\rm s}}\right\} \label{vp} .
\end{eqnarray}
The rotational velocity is slightly lower than the Keplerian velocity because the disk is supported against the gravity in part by the gas pressure. 
%and is 1.27~km~s$^{-1}$ at the outer edge, $ R _{\rm d} = 250~\mbox{au}$.
The disk mass is $ M _{\rm d} = 1.6\times10^{-3} M _\odot $.

\subsubsection{Cloudlet}\label{cloud}
We consider an ellipsoidal cloudlet accreting onto the protoplanetary disk though a warm neutral medium. 
The initial position of its center is located at $(x,y,z)=(5r_0,0,0)$. 
The surface of the cloudlet is defined as

\begin{eqnarray}
\frac{(x-5r_0)^2}{a_x^2}+\frac{y^2}{a_y^2}+\frac{z^2}{a_z^2}=1, 
\label{surface}
\end{eqnarray}
where $a_x,a_y$ and $a_z$ are half-lengths of the principal axes in the $x,y$ and $z$ directions, respectively.

The cloudlet has a uniform temperature of $T_{{\rm c}}=37~\rm K$ if the mean molecular weight is set to $\mu_{{\rm c}} = 2.3$. The cloudlet is in the pressure balance with the warm neutral medium.
The pressure and density distributions inside the cloudlet are described as
\begin{align}
p(r,z) &=p_0 \exp\left(-\frac{\mu _0 m _{\rm H} \Phi(r,z)}{k _{\rm B} T_0}\right),\\
\rho(r,z) & = \frac{\mu_{\rm c} m_{\rm H}}{k_{\rm B}T_{\rm c}}p(r,z) =  \frac{\mu_{\rm c} T_{0}}{\mu_{0} T_{\rm c}}\rho_0\exp\left(-\frac{\mu _0 m _{\rm H} \Phi(r,z)}{k _{\rm B} T_0}\right).
\end{align}

The initial cloudlet mass, $M_{\rm c}$ is listed in Table \ref{tab:table1}.

The velocity distribution inside the cloudlet is determined as follows. The initial velocity of the cloudlet is set by assuming that the orbit is parabolic with a periastron distance, $d = 100~\rm au $ (Figure \ref{schematic view}). The cloudlet has a uniform specific angular momentum $l_z=\sqrt{2GMd}=4.5\times 10^{20}~{\rm cm^2~s^{-1}}$ such that the pericenter is $d$. This requirement determines the velocity distribution of the cloudlet $(v_r,v_\varphi,v_z)$ as

\begin{eqnarray*}
\left( v _r, v _\varphi, v _z \right) & = &
\left[ - \frac{\sqrt{2GM (r - d)}}{r}, \frac{\sqrt{2GMd}}{r}, 0 \right].
\end{eqnarray*}

The initial magnetic field is assumed to be

\begin{eqnarray}
\left( B _x, B _y, B _z \right) & = & 
\begin{dcases}
\left( 0, 0, B _0 \right) & \mbox{if}~\frac{(x-5r_0)^2}{a_x^2}+\frac{y^2}{a_y^2} \le 1 \\
0 & \mbox{otherwise}
\end{dcases} 
,
\end{eqnarray}
where $ B _0 $ denotes the initial magnetic field strength. Only the cloudlet is threaded by a straight magnetic field (parallel to the $z$ axis), and the disk is unmagnetized (Figure \ref{schematic view}).
If we initially had imposed a magnetic field to the disk, the disk structure would have been largely affected by the magnetic field via e.g. magneto-rotational instability \citep[e.g.][]{Velikhov1959StabilityOA,1991ApJ...376..214B}. As we wish to focus on the dynamical interaction between the infalling cloudlet and the disk, we ignore the disk magnetic field to simplify the situation. The plasma $\beta$ of the cloudlet is $\sim 10$ so that the magnetic pressure is minor in and around the cloudlet. Therefore, the cloudlet expansion due to the magnetic pressure is insignificant.

We conducted simulations for eleven cloudlet models of various sizes as summarized in Table \ref{tab:table1}. The model name consists of the half size of the cloudlet, $a_z$, and the initial field strength. The values of $a_x$ and $a_y$ are fixed to 150 au. As shown later, the relative size between the cloudlet and the disk thickness, $a_z/z_{\rm s,max}$ is a crucial parameter for the resulting rotational velocity profile.

\begin{deluxetable*}{cccccccc}
%\tablenum{1}
\caption{The model parameters.}
\label{tab:table1}
\tablewidth{0pt}
\tablehead{
\colhead{Model} & \colhead{$L_x$} & \colhead{$L_y$} & \colhead{$L_z$} & \colhead{$a_z$} & \colhead{$a_z/z_{{\rm s,max}}$} & \colhead{$M_{\rm c}$} & \colhead{$B_0$}\\
\colhead{} & \colhead{$(\rm au)$} & \colhead{$(\rm au)$} & \colhead{$(\rm au)$} & \colhead{$(\rm au)$} & \colhead{} & \colhead{$(10^{-4}M_{\odot})$} & \colhead{$(\rm \mu G)$}
}
%\decimalcolnumbers
\startdata
    S60\_NB & 700 & 700 & 350 & 60 & 0.91 & 0.53 & 0\\
    S150\_NB & 1400 & 1400 & 350 & 150 & 2.27 & 1.3 & 0\\
    S60\_B58 & 700 & 700 & 350 & 60 & 0.91 & 0.53 & 58\\
    S150\_B58 & 1400 & 1400 & 350 & 150 & 2.27 & 1.3 & 58\\
    S60\_B82 & 700 & 700 & 350 & 60 & 0.91 & 0.53 & 82\\
    S150\_B82 & 1400 & 1400 & 350 & 150 & 2.27 & 1.3 & 82\\
    S60\_B116 & 700 & 700 & 350 & 60 & 0.91 & 0.53 & 116\\
    S90\_B116 & 700 & 700 & 350 & 90 & 1.36 & 0.80 & 116\\
    S120\_B116 & 700 & 700 & 350 & 120 & 1.82 & 1.1 & 116\\
    S150\_B116 & 700 & 700 & 350 & 150 & 2.27 & 1.3 & 116\\
    S180\_B116 & 700 & 700 & 420 & 180 & 2.73 & 1.6 & 116
\enddata
\tablecomments{$a_z/z_{\rm s,max}$ denotes the relative size between the cloudlet and the disk thickness.}
\end{deluxetable*}

We describe two major limitations: the timescale and hysteresis of our models. 
The dynamical interaction between the cloudlet and the disk occurs on a much shorter timescale ($\rm \sim 10^3 ~yr$) than the viscous timescale ($\rm \sim 10^6 ~yr$). Considering this, we ignore the details associated with disk accretion, such as an effective viscosity and a disk magnetic field. Our models are therefore incapable of studying the long-term evolution. In addition, our models focus on the single cloudlet encounter event, though cloudlet capture events are likely to occur repeatedly during the viscous timescale (see, e.g., \cite{kuffmeier2018}). Our model cannot deal with the situations where the hysteresis of previous events is significant.

\subsubsection{Numerical Methods}
We solved Equations (\ref{mass}) through (\ref{pressure}) using CANS+ \citep{matsumoto2019PASJ}.
The basic equations are solved in the Cartesian coordinates, $ (x, y, z) $. 
We adopt the Harten--Lax--van Leer Discontinuities (HLLD) approximate Riemann solver
of \cite{miyoshi&kusano2005} and the hyperbolic divergence cleaning method \citep{dedner2002JCoPh}. We employ MP5 \citep{Suresh&Huynh1997} and a third-order Runge-Kutta method to achieve the fifth-order accuracy in space and third-order accuracy in time.

The computation domain is a rectangular box that covers $|x|<L_{x}$, $|y|<L_{y}$ and $|z|<L_{z}$, where $L_{x},~L_{y}$, and $~L_{z}$ take different values depending on the models. The values of $L_{x},~L_{y}$ and $L_{z}$ are listed in Table~\ref{tab:table1}. We fix the spatial resolution for all the models, and $\Delta x =\Delta y = \Delta z=7$ au. For simplicity, we apply the fixed boundary conditions to all the physical variables ($\rho$, $p$, $\bm{v}$, $\bm{B}$) in all the directions. The fixed boundary condition for the magnetic field may not be natural. We discuss the effects of the fixed boundary condition in Appendix \ref{boundary}.

\section{Result} \label{Result}

We first present the results of two models with and without magnetic fields, S60\_B116 and S60\_NB, respectively, to highlight the importance of magnetic fields. We then show how the impact of magnetic fields depends on the cloudlet size and the cloudlet field strength.

\subsection{Models with and without magnetic fields}\label{subsec:E1NB&E1}

We compare the results of Models S60\_B116 and S60\_NB, which only differ in the presence of magnetic fields. In both models, the relative size of the cloudlet to the disk thickness is close to unity (0.91). The initial plasma $\beta$ at the center of the cloudlet of S60\_B116 is 14, where the plasma $\beta$ is defined as the ratio of the gas pressure to the magnetic pressure. The field strength is $B_0 = 116~{\rm \mu G}$.

\begin{figure*}
\centering
\plottwo{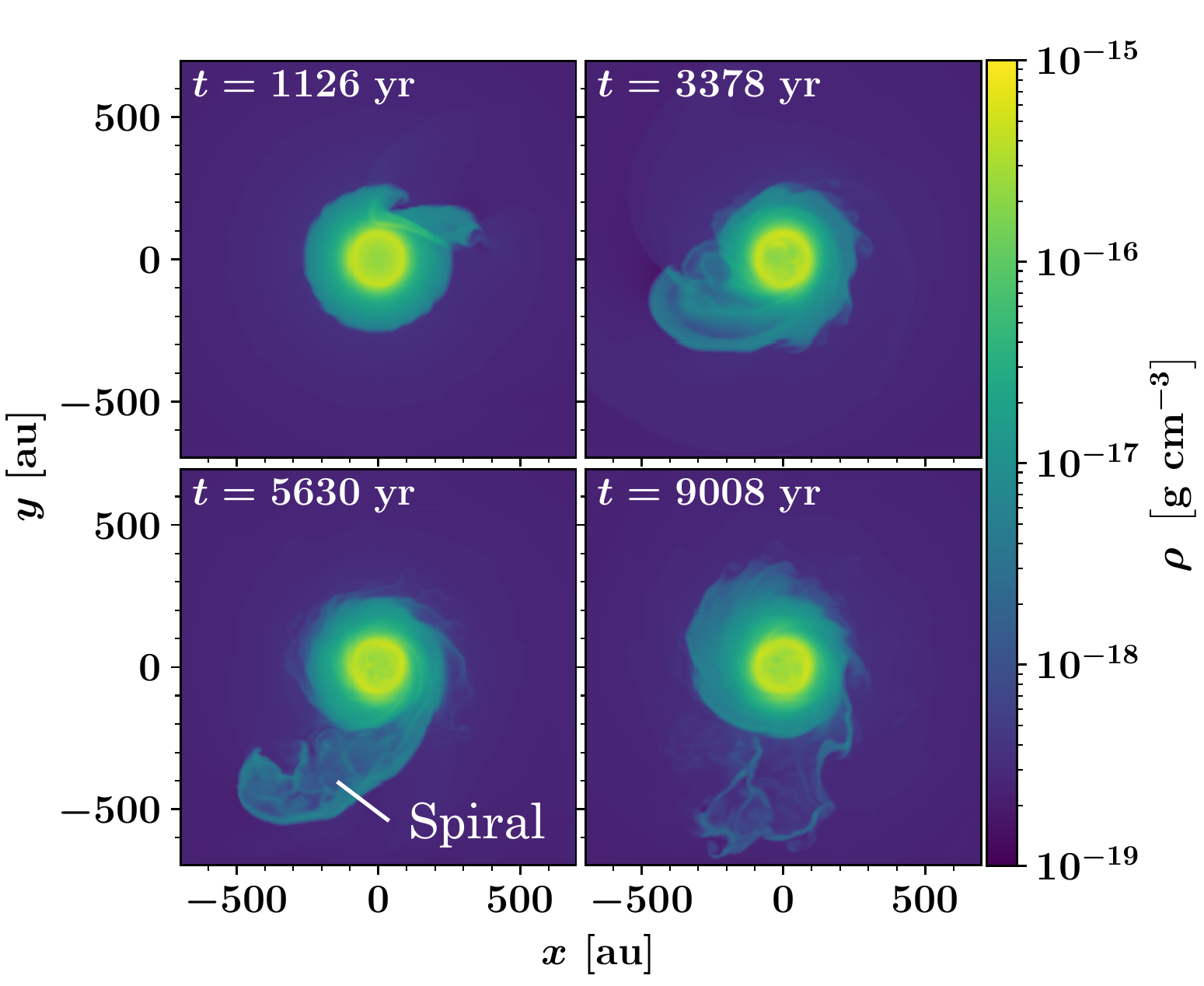}{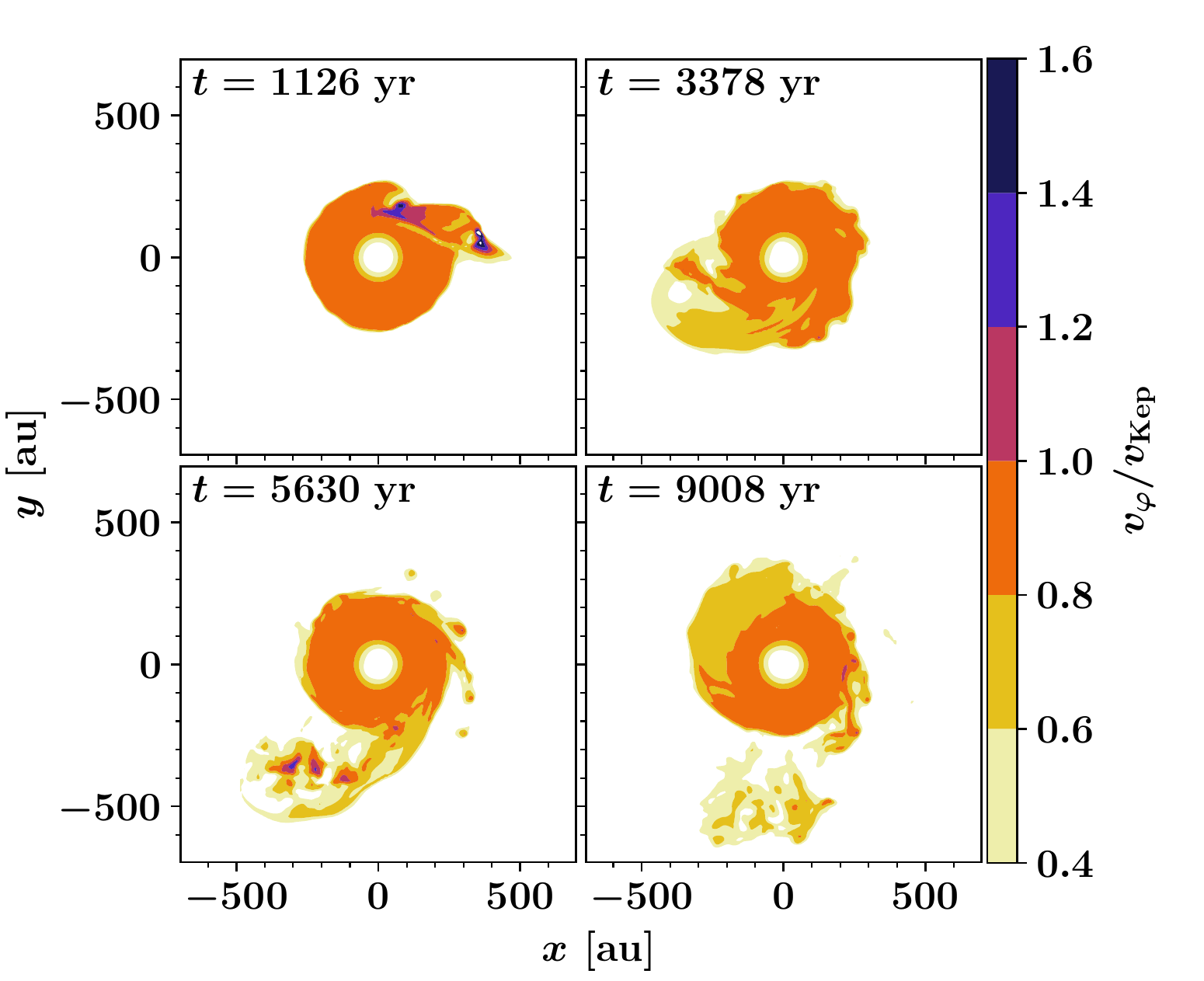}
\caption{Evolution of the density (left) and the rotational velocity (right) distributions at the midplane $z=3.5~{\rm au}$ for the hydrodynamic model, S60\_NB. The rotational velocity is normalized by the local Keplerian orbital velocity, $v_{\rm Kep}$.
\label{fig:e1nb}}
\end{figure*}

Figure \ref{fig:e1nb} displays the evolution of the density (left) and rotational velocity (right) distributions on the midplane for the hydrodynamic model, S60\_NB. The rotational velocity is normalized by the local Keplerian orbital velocity, $v_{\rm Kep}=\sqrt{GM/r}$. At $t = 1126~\rm yr$, an outer part of the disk is broken by the cloudlet. After the encounter, a part of the colliding cloudlet material is ejected away from the disk, forming a spiral structure as shown in the panels of $t = 3378~\rm yr$ and $5630~\rm yr$. A large fraction of the gas in the spiral falls back onto the disk. A diffuse remnant of the ejected material remains outside the disk. The rotational velocity is not higher than the local Keplerian orbital velocity except for a narrow region of the cloudlet impact in an early stage ($t = 1126~\rm yr$).

\begin{figure*}
\begin{center}
\centering
\plottwo{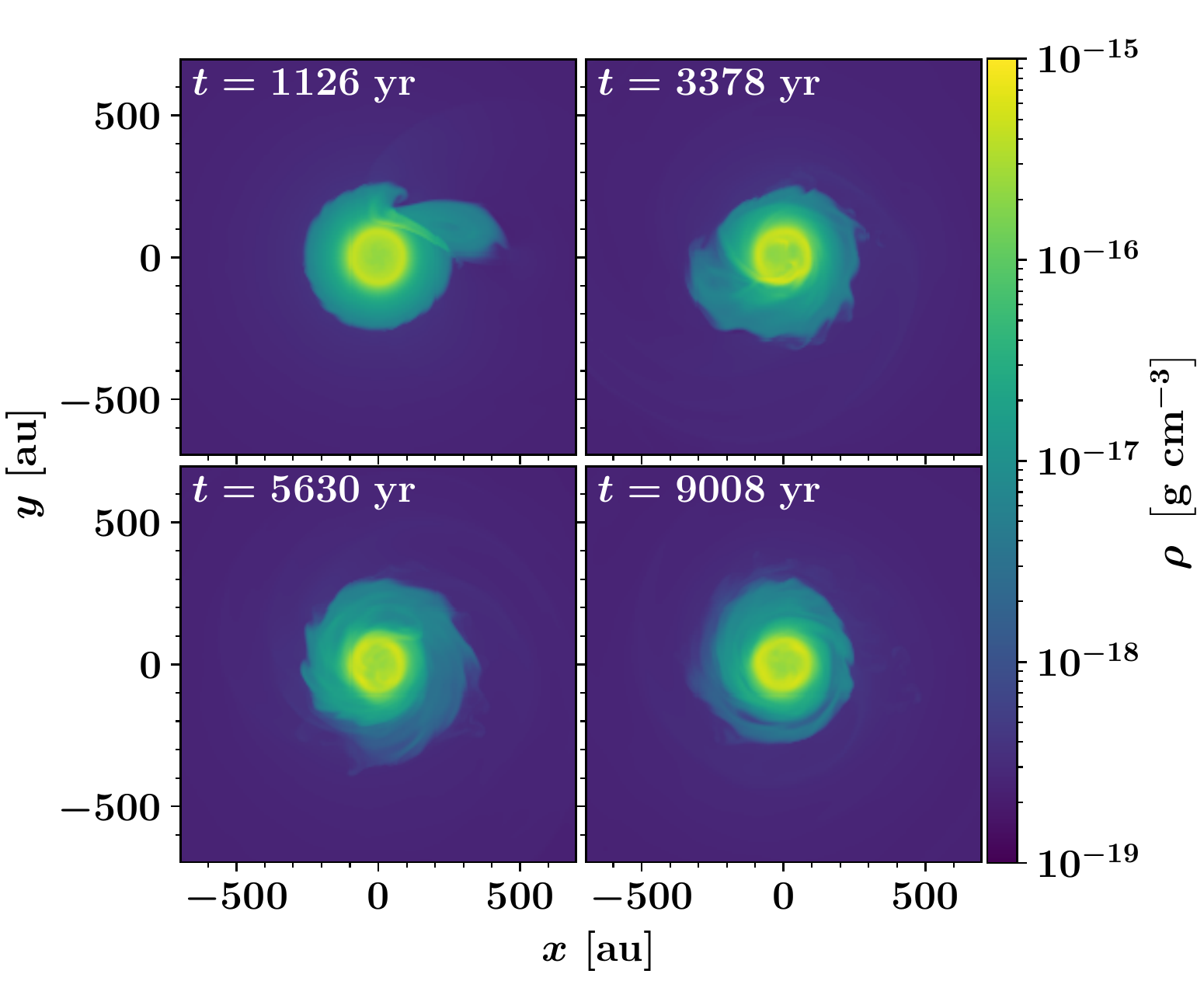}{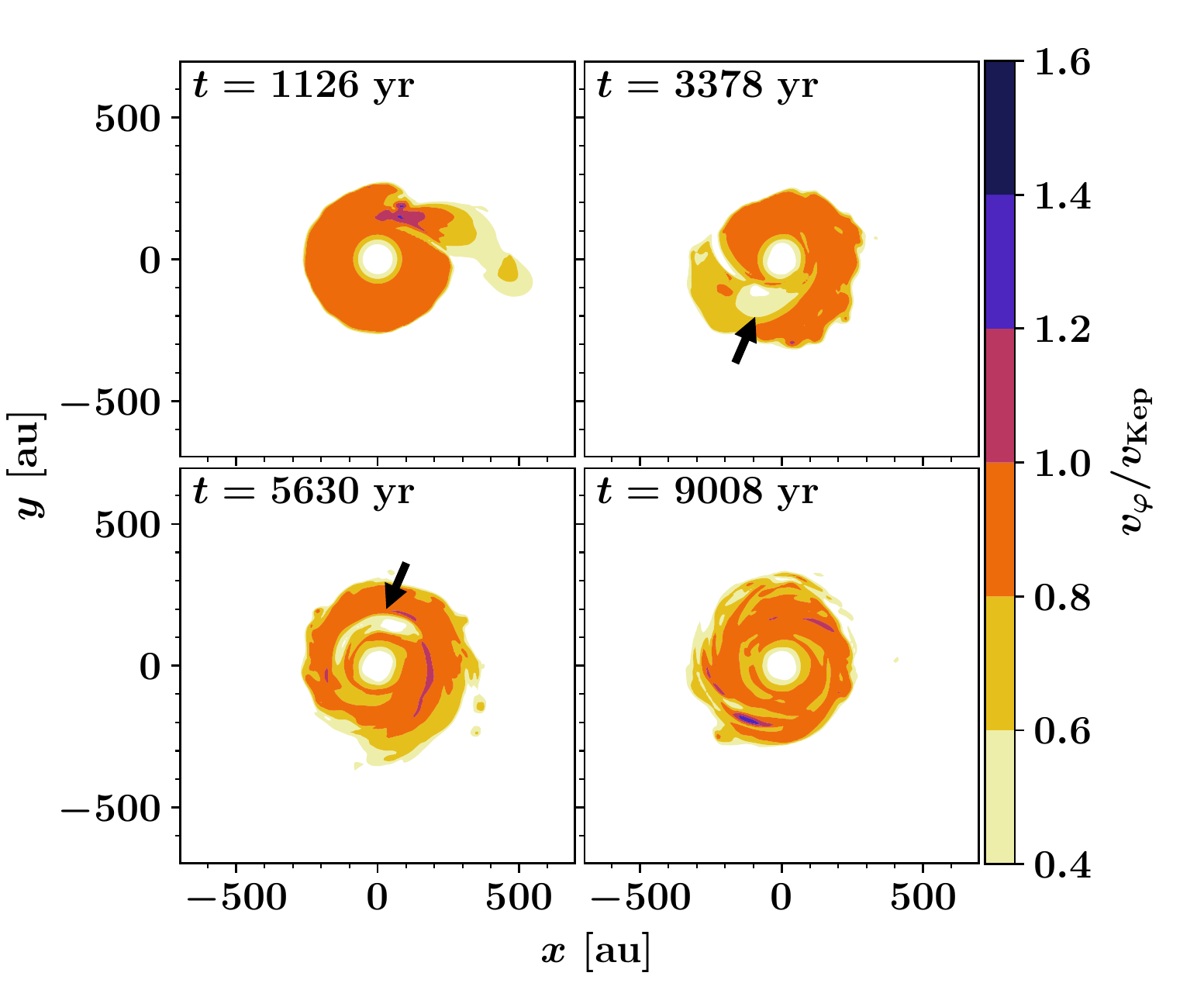}
\caption{
Same as Figure \ref{fig:e1nb}, but for Model S60\_B116. Black arrows indicate a local reduction in the rotational velocity. \label{fig:e1}
}
\end{center}
\end{figure*}

Figure \ref{fig:e1} is the same as Figure \ref{fig:e1nb}; however, for model S60\_B116, the spiral arm forms but does not grow in size in model S60\_B116. The growth is likely to be suppressed by the magnetic tension force acting on the cloudlet. Gas ejection is not appreciable, and accordingly, the fallback of the ejecta is insignificant. Instead, there appears an arc-like region where the rotational velocity is half of the local Keplerian orbital velocity at $t = 5630~\rm yr$. The reduction is a direct consequence of magnetic braking. The arc is wound up to be narrower at $t = 9008~\rm yr$. When the field strength is weak ($B_0 = \rm 58~\mu G$ (S60\_B58) and $B_0 = \rm 82~\mu G$ (S60\_B82)), the results are almost the same with S60\_B116 although the rotational velocity reduction is small.

\begin{figure*}

\plotone{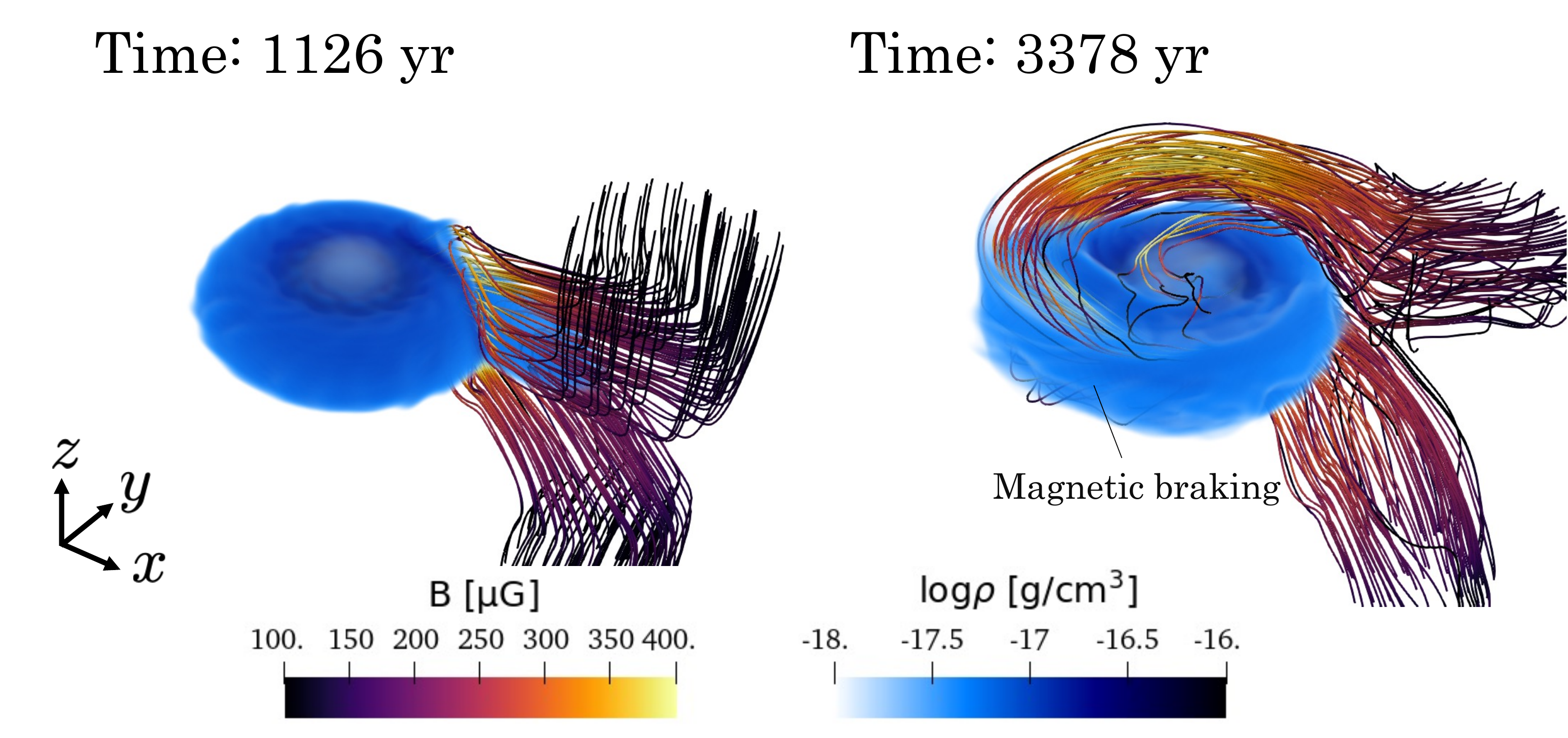}

\caption{
3D structure of Model S60\_B116. The gas distribution is shown by the volume rendering. Lines denote magnetic field lines. The color of field lines indicates the field strength. In the animation, the sequence starts at time 0 yr and ends at time 11260 yr. The video duration is 13 s. (An animation of this figure is available.)
}\label{fig:e13d}
\end{figure*}

We visualized the three-dimensional structure of magnetic fields in Figure \ref{fig:e13d}. Line colors denote the field strength. We also show the mass density distribution by volume rendering. Magnetic fields are dragged and stretched by the infalling cloudlet. As a result, magnetic tension force decelerates the rotational motion of the colliding cloudlet material.

\subsection{Dependence on the cloudlet size} \label{subsec:Models of different cloudlet thickness}
We investigate the dependence of the resulting rotational velocity profile on the cloudlet size for the models with magnetic fields.

Here, we present the results of Model S150\_B116 as an example to highlight the importance of the relative size of the cloudlet to the disk thickness. In this model, the vertical cloudlet size is approximately twice as large as the disk thickness. Therefore, the top and bottom parts of the cloudlet do not collide with the disk material at the time of impact and slide on the disk surfaces. We will show that magnetic fields {\it accelerate} the rotational motion in such cases.

\begin{figure*}
\plottwo{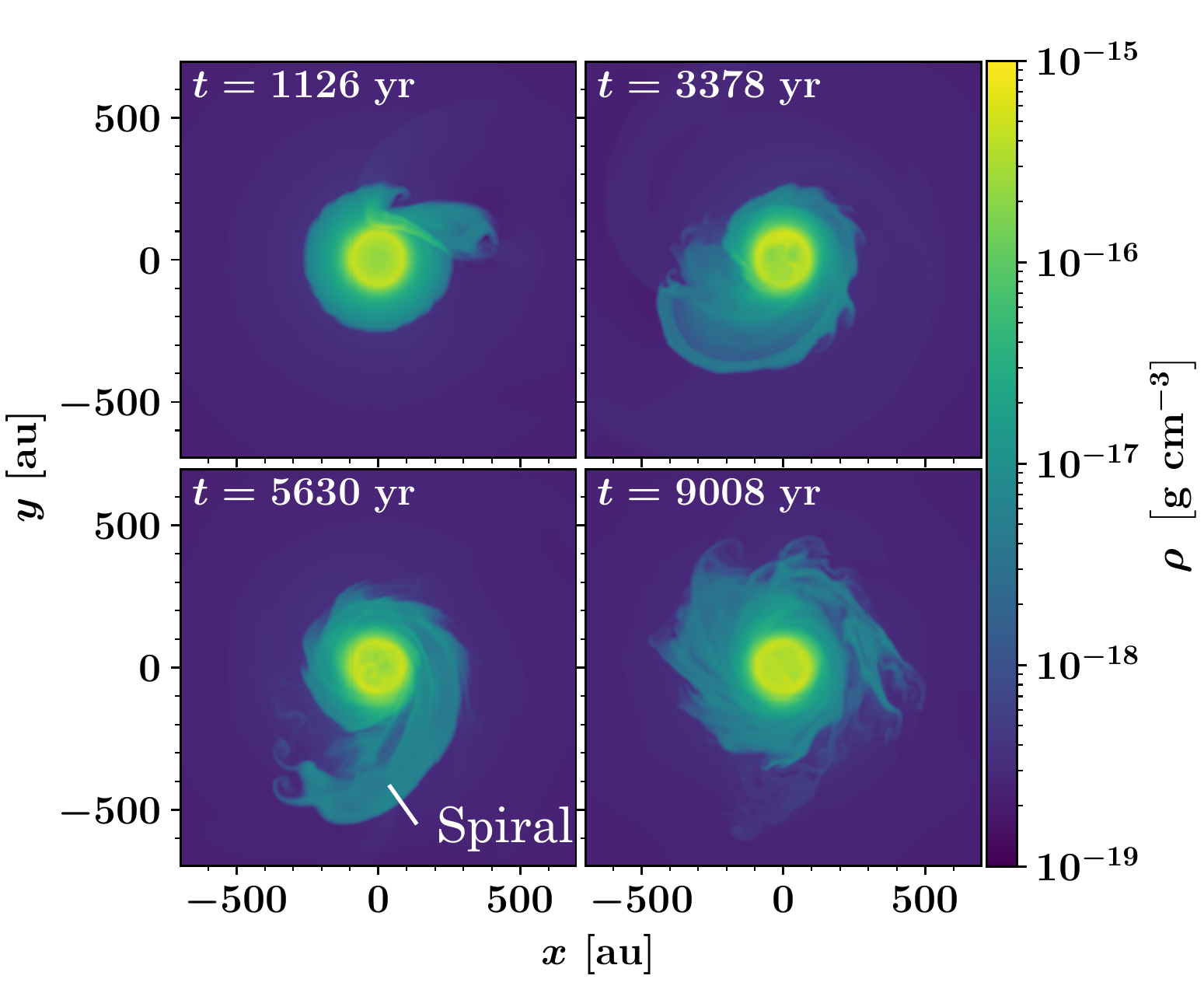}{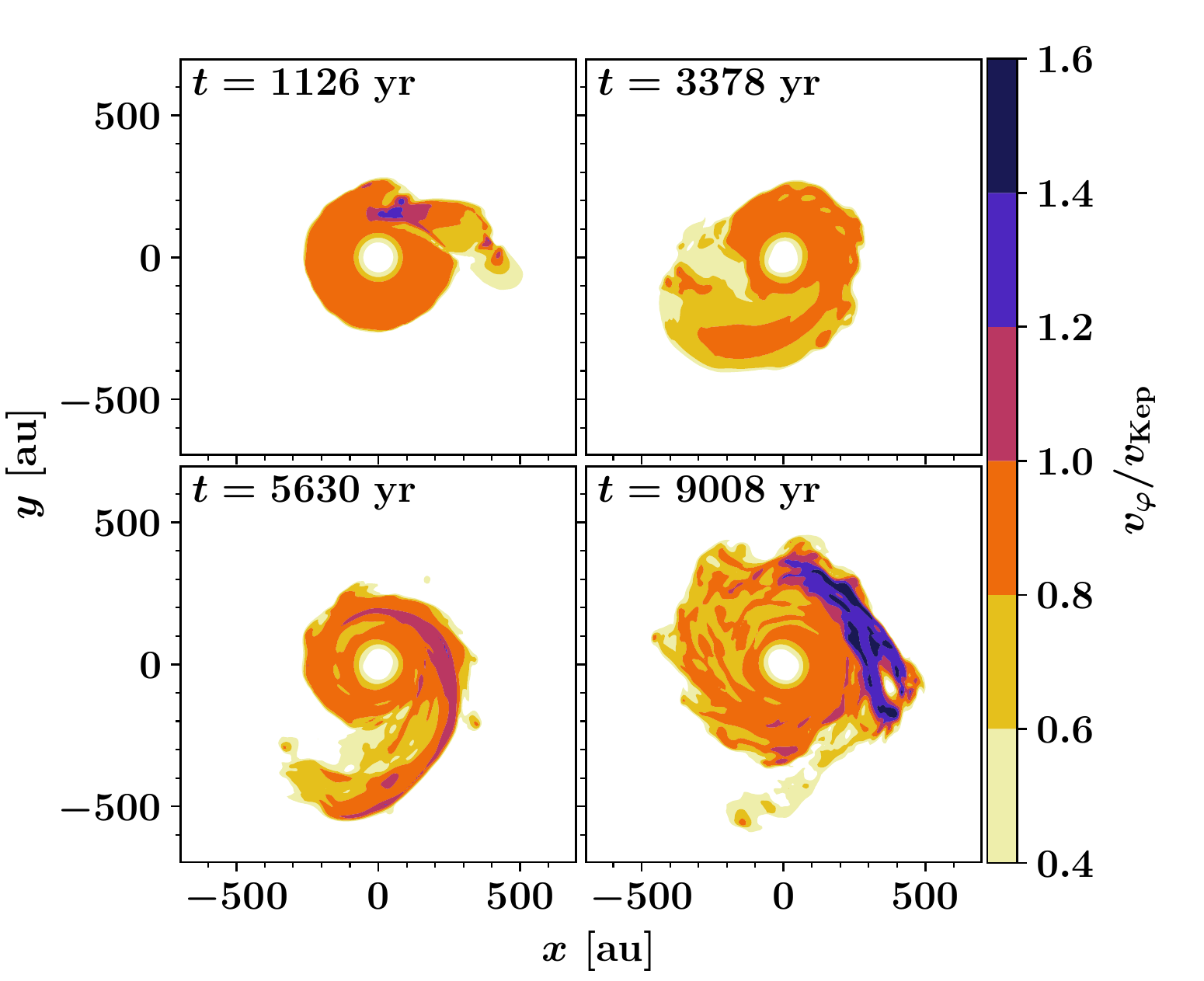}
\caption{
Same as Figure \ref{fig:e1nb}, but for Model S150\_B116.
\label{fig:s2}
}
\end{figure*}

Figure \ref{fig:s2} displays the midplane density (left) and rotational velocity (right) distributions at different times for Model S150\_B116. A spiral structure is formed even in the presence of magnetic fields, suggesting magnetic acceleration. The rotational velocity maps indicate that the colliding cloudlet material is accelerated to a super-Keplerian velocity. A fraction of the cloudlet becomes gravitationally unbound as a result of magnetic acceleration. 

\begin{figure*}

\plotone{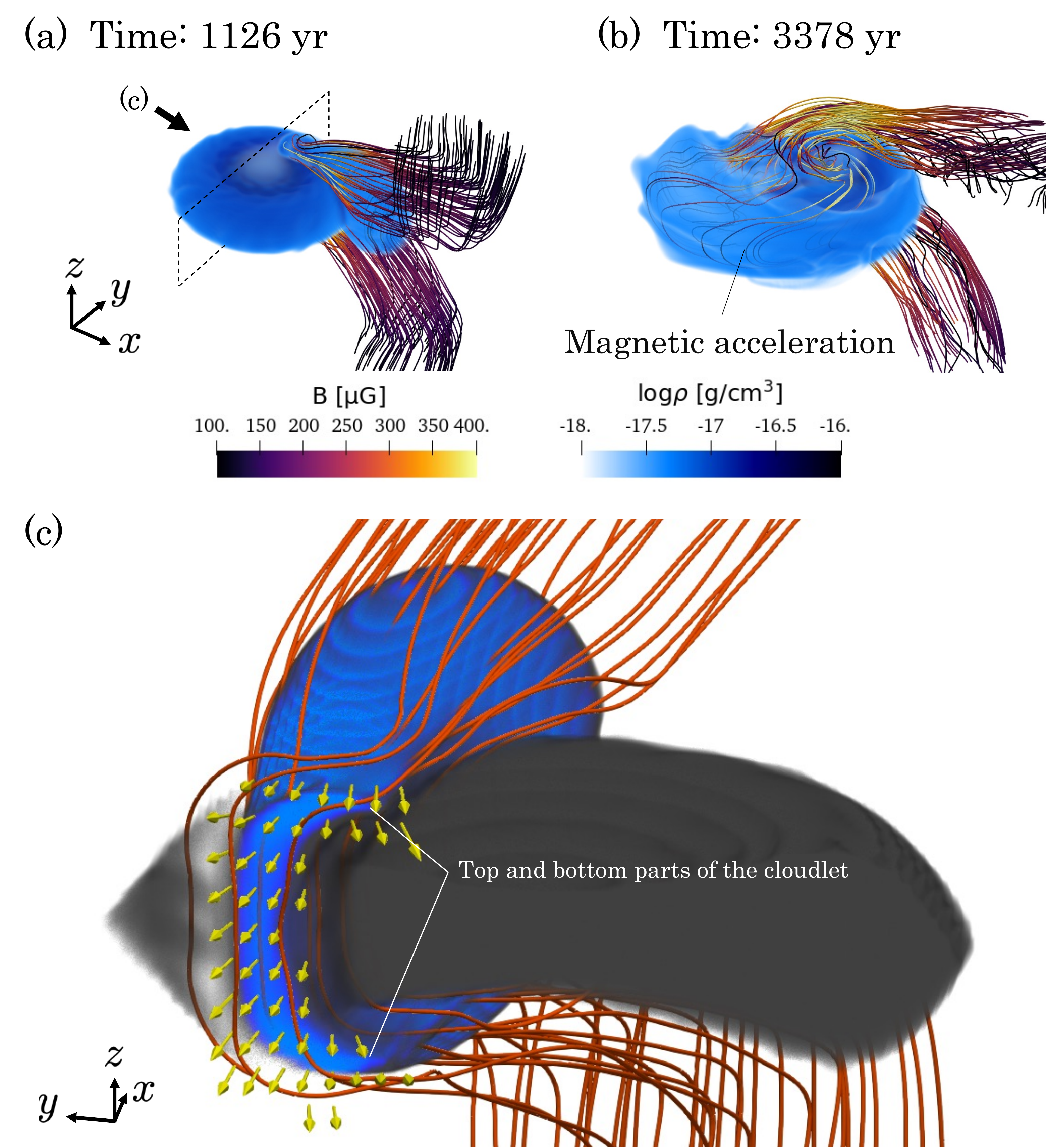}

\caption{Panels (a) and (b) show the volume rendering images of the mass density and magnetic field lines for Model S150\_B116 at two different times. Panel (c) is an enlargement of panel (a) but from a different viewing angle, which is shown by the black arrow in panel (a). The cloudlet and disk are colored in blue ($35\sim45~\rm K$) and grey ($45\sim55~\rm K$), respectively.  Vector arrows show the normalized velocity of the cloudlet on the slice at $x = 3.5~\rm au$. Orange solid lines indicate magnetic field lines. The top and bottom parts of the cloudlet slide on the disk surface against the colliding cloudlet material. The vertical velocity shear results in the magnetic field structure accelerate the colliding cloudlet material by magnetic tension force. In the animation, the sequence starts at time 0 yr and ends at time 11260 yr. The video duration is 13 s. (An animation of this figure is available.)
}\label{fig:s23d}
\end{figure*}

\begin{figure*}
\plotone{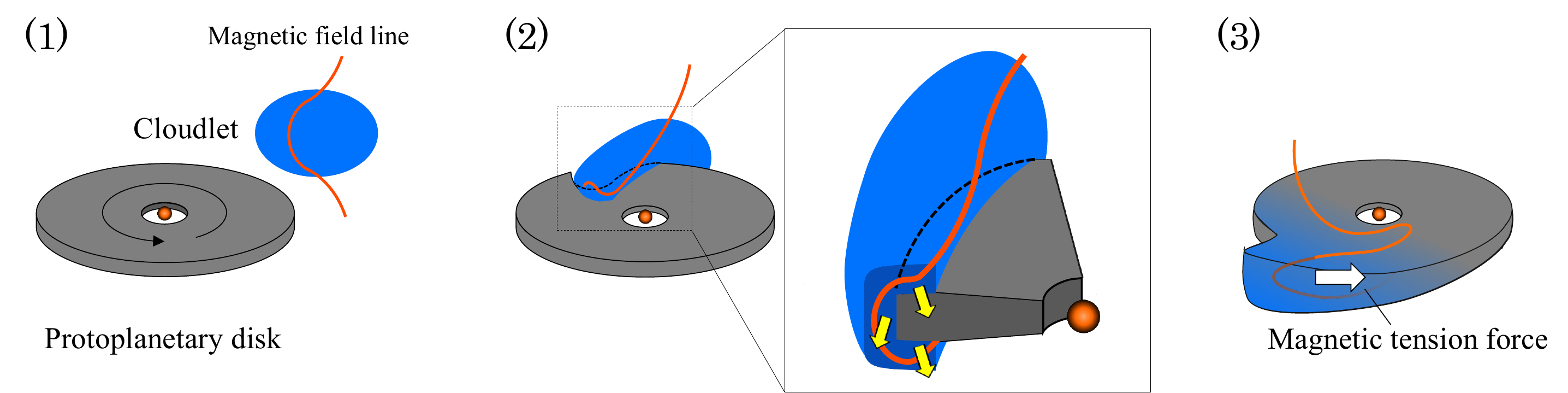}
\caption{Schematic diagram of magnetic acceleration. Panel (1) shows the phase before the cloudlet collision. Panel (2) displays the cloudlet collision, corresponding to Panel (a) of Figure \ref{fig:s23d}. The yellow arrows correspond to the vector arrows in Panel (c) of Figure \ref{fig:s23d}. Panel (3) shows magnetic acceleration, corresponding to Panel (b) of Figure \ref{fig:s23d}. 
\label{fig:MRA}}
\end{figure*}

Figure \ref{fig:s23d} describes the 3D structure of magnetic fields that accelerate the colliding cloudlet material. Panels (a) and (b) show the birds-eye-view images of the density structure at two different times. As in the case of Model S60\_B116, the magnetic fields of the cloudlet are highly stretched and amplified. However, the magnetic field geometry is different between the two models; in Model S150\_B116, the magnetic tension force is operating to accelerate the disk rotational motion while in Model S60\_B116 it decelerates the region. Panel (c) is an enlargement of Panel (a) but from a different viewing angle, which is shown by the black arrow in panel (a). The cloudlet and disk are colored in blue ($35\sim45~\rm K$) and grey ($45\sim55~\rm K$), respectively. Vector arrows show the normalized velocity of cloudlet on the slice at $x = 3.5~\rm au$. After the cloudlet encounters the disk, the cloudlet material around the midplane is shock-compressed and decelerates in the radial direction. However, the top and bottom parts of the cloudlet do not collide with the disk body. Instead, they slide on the disk surfaces along the parabolic orbit (see panel (c) of Figure \ref{fig:s23d}). Namely, the top and bottom parts retain a significant radial component of the velocity. As a result, magnetic fields are transferred toward the center more quickly around the disk surfaces than around the disk midplane. The flows sliding on the disk surfaces move toward the center to spin up, twisting up magnetic fields. The vertical velocity shear results in a magnetic field structure that accelerates the colliding cloudlet material through the magnetic tension force. This example demonstrates the importance of the relative size of the cloudlet to the disk thickness. Figure \ref{fig:MRA} shows a schematic diagram of this process. For comparison, we also investigated the non-magnetized model (S150\_NB). We confirmed that the model has almost the same result as S60\_NB; a part of colliding cloudlet material is ejected and forms a sub-Keplerian spiral structure. We will summarize the magnetic and non-magnetic cases in Section \ref{sec:summary}.

To demonstrate that magnetic tension force indeed accelerates the colliding cloudlet material on a timescale $\sim 10^3~\rm yr$, which is shorter than the Keplerian orbital period, we evaluated $\tau_{\rm mag}$, the timescale of acceleration by magnetic tension force required for the gas to reach the local escape velocity. $\tau_{\rm mag}$ is given by the work rate of magnetic tension force $(v_r\bm{e_r}+v_\varphi\bm{e_\varphi})\cdot (\left(\bm{B}\cdot\bm{\nabla}\right)\bm{B})/4\pi$ and expressed as,
\begin{align}
    \tau_{\rm mag} = \frac{GM\rho/(r^2+z^2)^{1/2}}{(v_r\bm{e_r}+v_\varphi\bm{e_\varphi})\cdot (\left(\bm{B}\cdot\bm{\nabla}\right)\bm{B})/4\pi}.
\end{align}
$\tau_{\rm mag}$ can take either the positive and negative signs, and the positive and negative signs indicate the acceleration and deceleration timescales, respectively.

\begin{figure*}
\plotone{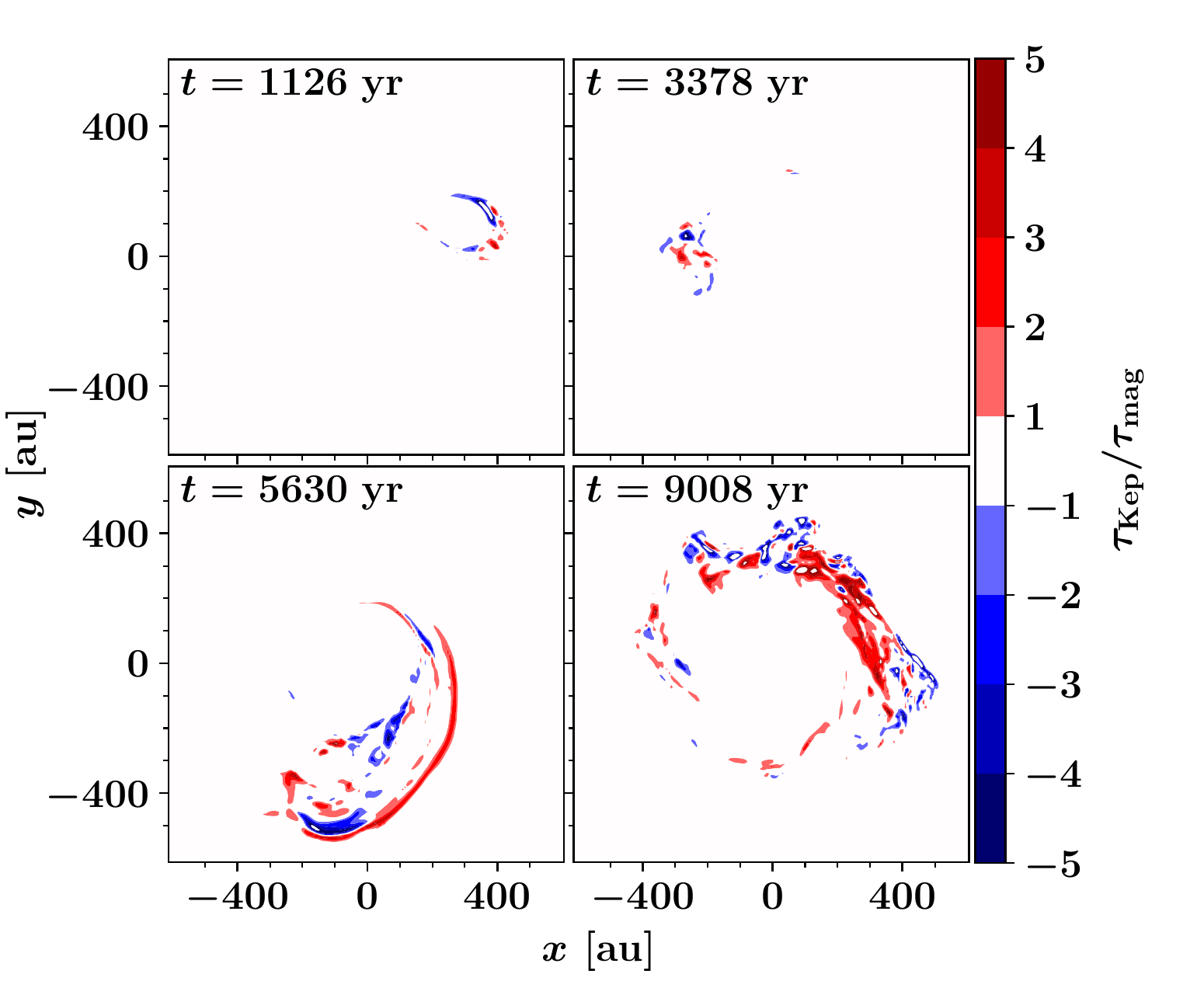}
\caption{The ratio of Keplerian orbital period $\tau_{\rm Kep}$ to acceleration timescale by magnetic tension force $\tau_{\rm mag}$ at the midplane for Model S150\_B116. $\tau_{\rm Kep}/\tau_{\rm mag}$ is larger than two in the spiral and super-Keplerian regions where $\tau _{\rm Kep} $ = $7.4 \times 10 ^3 $ yr and $1.6 \times 10 ^4$ yr at $r = 300 \rm ~au$ and 500 au, respectively. It means that the acceleration timescale is $\tau_{\rm mag} \leq 3.7\times10^3$ and $8.0\times10^3~\rm yr$ at $r = 300 \rm ~au$ and 500 au, respectively. The acceleration can be explained by magentic tension force.
}
\label{fig:tension}
\end{figure*}

Figure \ref{fig:tension} displays the midplane distributions of $\tau_{\rm Kep}/\tau_{\rm mag}$ at different times, where $\tau_{\rm Kep}=2\pi(r^3/GM)^{1/2}$ is the Keplerian orbital period. 
The bottom panels show that $\tau_{\rm Kep}/\tau_{\rm mag}$ is larger than two in the spiral and super-Keplerian regions where $\tau _{\rm Kep} $ = $7.4 \times 10 ^3 $ yr and $1.6 \times 10 ^4$ yr at $r = 300 \rm ~au$ and 500 au, respectively. It means that the acceleration timescale is $\tau_{\rm mag} \leq 3.7\times10^3$ and $8.0\times10^3~\rm yr$ at $r = 300 \rm ~au$ and 500 au, respectively. The acceleration timescale can be shorter if we take account of the preexisting kinetic energy before magnetic acceleration. Therefore, we confirmed the rapid acceleration by magnetic tension force.

\subsection{Dependence of the mass of the gravitationally unbound gas on the cloudlet size} \label{unbound}
We have shown that magnetic fields can either decelerate or accelerate the rotational motion of the colliding cloudlet material, depending on the relative size of the cloudlet to the disk thickness. When the cloudlet size is larger than the disk thickness, the vertical velocity shear in the colliding cloudlet can result in the acceleration of rotational motion. As a result, a fraction of accelerated gas becomes gravitationally unbound.

\begin{figure}
\plotone{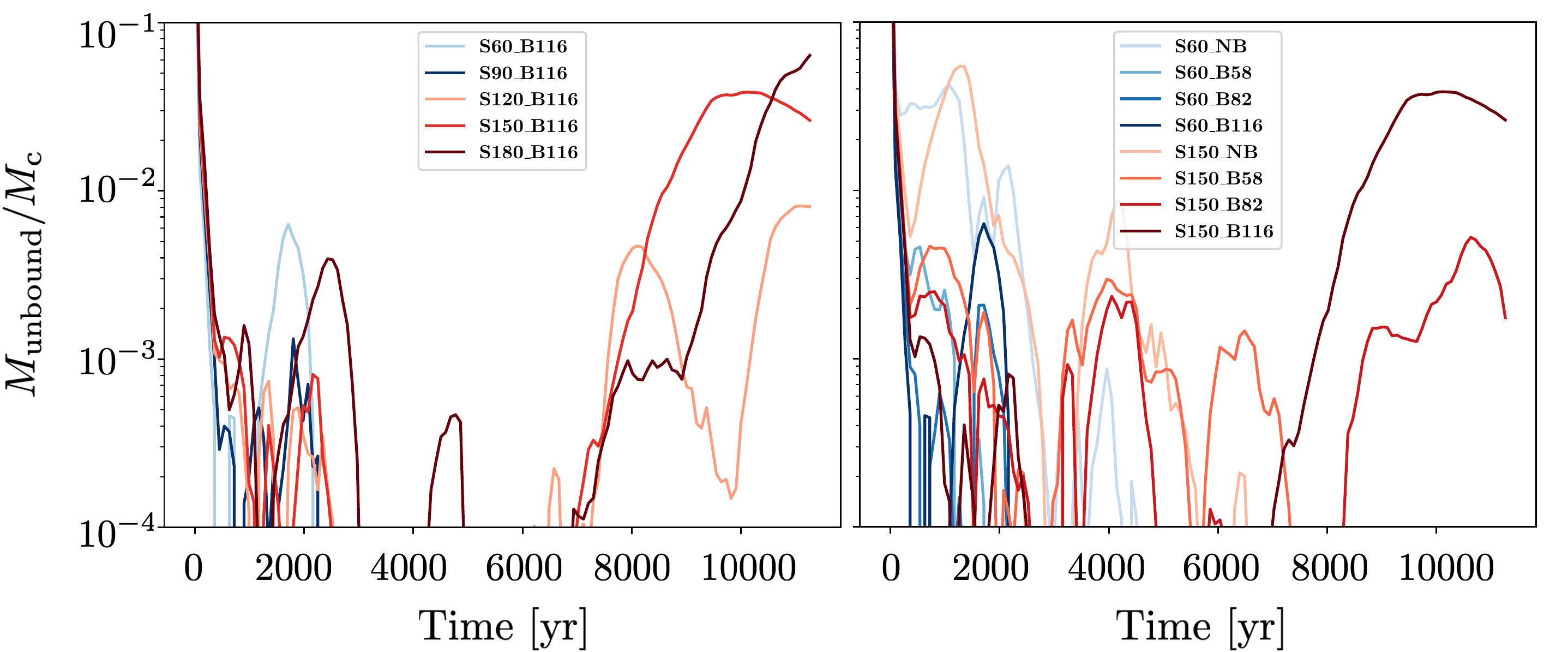}
\caption{Time evolution of the ratio of the mass of the gravitationally unbound gas to the initial cloudlet mass for all models.
\label{fig:unbound_mass}}
\end{figure}

To evaluate the size dependence of the mass of the gravitationally unbound gas $M_{\rm unbound}$, we investigated the time evolution of the mass of the gravitationally unbound gas in all models. The results are shown in Figure \ref{fig:unbound_mass}, where the mass is normalized by the initial cloudlet mass $M_{\rm c}$. The left panel shows the dependence of cloudlet size. The right panel shows the dependence of magnetic field strength. The mass of the gravitationally unbound gas is evaluated in the regions with the conditions of $\rho\geq10^{-18}~\rm g~cm^{-3}$ and $|z|\le h$, where the definition of $h$ depends on the relative size of the cloudlet:
\begin{align}
h = 
\begin{dcases}
z_{\rm s,max}&~\left(\mathrm{if~}a_z < z_{\rm s,max}\right), \\
a_z&~\left(\mathrm{if~}a_z > z_{\rm s,max}\right).\nonumber
\end{dcases}
\label{g-star}
\end{align}
Approximately 1--10\% of the initial cloudlet mass is found to be gravitationally unbound in Models S120\_B116, S150\_B116, and S180\_B116. In these models, the cloudlet sizes are larger by a factor of $\sim$2 to 3 when compared to the disk thickness, and the field strengths are larger than $1\times10^2$ $\mu$G. In the other models, the mass of the gravitationally unbound gas is negligibly small. The above results provide constraints on the relative size of the cloudlet and the field strength of magnetic acceleration.

\section{discussion}\label{sec:discussion}
We performed 3D MHD simulations with different cloudlet sizes and magnetic field strengths.
The simulations demonstrate that the rotational velocity of the colliding gas can either be sub-Keplerian or super-Keplerian, depending mainly on the relative size of the cloudlet to the disk thickness. When the cloudlet size is comparable to or smaller than the disk thickness, the rotation motion of the colliding cloudlet material is only decelerated by magnetic braking. However, if the cloudlet size is larger than the disk thickness, the colliding cloudlet material can rotate at a super-Keplerian velocity as a result of magnetic acceleration. We showed that the vertical velocity shear of the cloudlet produces a magnetic tension force that increases the rotational velocity (Figures~\ref{fig:s23d} and \ref{fig:MRA}). 

\begin{figure*}
\plotone{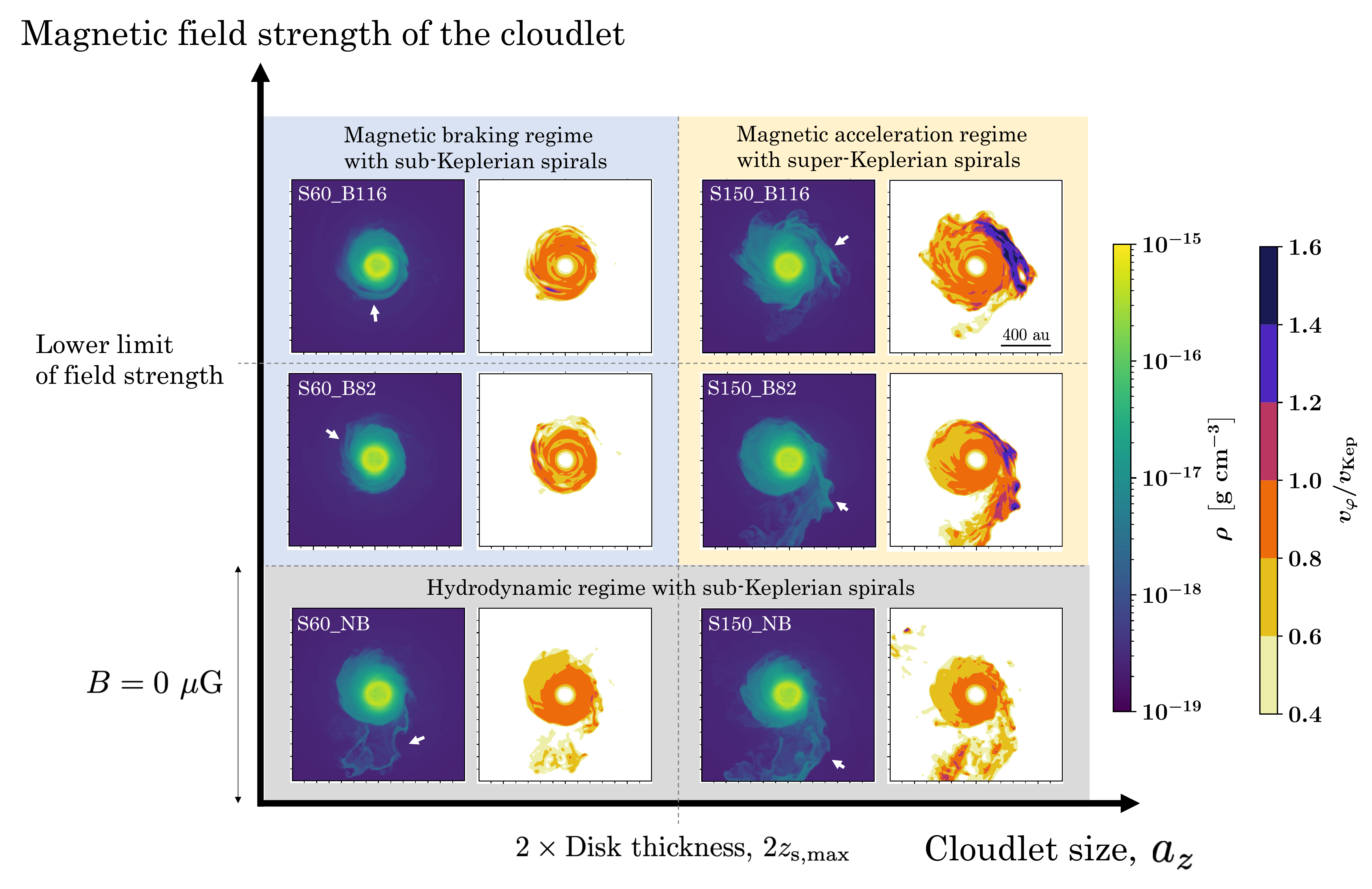}
\caption{Phase diagram under different conditions of cloudlet size ($a_z$) and magnetic field strength  ($B_0$) at $t = 9008~\rm yr$. White arrows indicate spirals.}
\label{fig:phase}
\end{figure*}

Our model shows that the spiral or arm can rotate at a super-Keplerian velocity as a result of magnetic acceleration (Figure \ref{fig:s2}).
Such a super-Keplerian (non-Keplerian) spiral structure is found around RU Lup, a class II object \citep[e.g.][]{alcala2017,andrew2018,Bailer-Jones2018,gaia2018,Yen2018}. According to \citet{Huang2020}, gravitationally unbound clumps are distributed along the spiral arms. Their mass is estimated to be $\sim 0.1-150~M_{\oplus}$. It has been considered that the clumpy spirals are results of gravitational instability \citep[e.g.][]{Durisen2007,Vorobyov2016}. However, our results suggest that such structures can also be formed through the cloudlet capture event and magnetic acceleration. In Section \ref{unbound}, we showed that 1--10\% of the initial cloudlet mass goes to the gravitationally unbound gas if magnetic acceleration efficiently operates (see also Figure \ref{fig:unbound_mass}). If the initial cloudlet mass is $\sim 1\times10^{-4}~M_\odot$, then the gravitationally unbound gas with the mass of $\sim 0.3 - 3~M_{\oplus}$ can be produced.

In the following, we discuss some conditions under which magnetic acceleration can be expected.

\subsection{Magnetic field strength required for magnetic acceleration}\label{critical}
We estimate the lower limit of the initial field strength $B_0$ of the cloudlet required for magnetic acceleration. The field strength of the cloudlet is mainly amplified after the encounter via shock compression. The field strength will be further increased by the vertical velocity shear (Figures \ref{fig:s23d} and \ref{fig:MRA}), however, we assume that the dominant amplification process is shock compression. For the amplified magnetic fields to produce gravitationally unbound structures, the magnetic energy density should be larger than the gravitational energy density at the accelerated region (spiral in Figure \ref{fig:s2}). Considering this, we require the relation
\begin{align}
\frac{B_{\rm post}^2}{8\pi} \ge \frac{GM\rho}{r},
\end{align}
where $B_{\rm post}$ is the field strength after shock compression, and the value is $B_0$ times the compression ratio. The compression ratio can be obtained from the Rankine-Hugoniot relation for perpendicular shocks (e.g., Equation (5.35) of \cite{2014priest}). For $r \sim 400~\rm au$ and $\rho\sim5\times10^{-18}~\rm g~cm^{-3}$, the lower limit is estimated as $B_0\sim1\times10^2~\rm \mu G$ when the cloudlet has $v_{\rm pre}\sim1.7~\rm km~s^{-1}$, $T_{\rm pre}\sim38~\rm K$, $\rho_{\rm pre}\sim1\times10^{-17}~\rm g~cm^{-3}$ just before the collision. The upper limit of the initial plasma $\beta$ at the center of the cloudlet is $\sim2\times10^1$.
The right panel of Figure \ref{fig:unbound_mass} supports the validity of the above estimation. Among the models S150\_B58, S150\_B82 and S150\_B116, only Model S150\_B116 has a field strength larger than the lower limit. The figure shows that Model S150\_B116 produces a significantly larger mass of unbound gas than the other two models. However, the magnetic field becomes strong enough before the collision of the cloudlet with the disk, magnetic braking should suppress the dynamical infall of the cloudlet. Therefore, the acceleration occurs when the field strength is stronger than the above lower limit but not strong enough to suppress the infall of cloudlet before the collision.

\subsection{Non-ideal MHD effect} \label{Non-ideal}
The protoplanetary disk is generally in a low ionization state \citep[e.g.][]{gammie1996}, which makes the non-ideal MHD effects important.
As magnetic diffusion results in a weak magnetic acceleration, we estimate the impact. We only consider ambipolar diffusion. In the plasma composed of ions, electrons, and neutrals, the ambipolar diffusion coefficient $\eta_{\rm AD}$ \citep[e.g.][]{2007Ap&SS.311...35W,2016A&A...587A..32M} is given by,

\begin{eqnarray}
\eta_{\rm AD} &=& \frac{B^2}{4\pi \gamma_{\rm AD}\rho_n \rho_i},\\
\gamma_{\rm AD} &=& \frac{\left<\sigma_{in} v_{i}\right>}{m_i + m_n},
\end{eqnarray}
where $\gamma_{\rm AD},~\left<\sigma_{in} v_{i}\right>,~\rho_n,~\rho_i,~m_n, m_i$ are the drag coefficient, the ion-neutral collision rate, the neutral density, the ion density, the mass of neutral particle and the mass of ion particle. The ion-neutral collision rate is $\left<\sigma_{in} v_{i}\right>\sim 2\times10^{-9}~\rm cm^{3}~s^{-1}$ \citep{1961ApJ...134..270O} for $m_n\sim2 m_{\rm H}$ and $m_i\sim10 m_{\rm H}$. We also introduce the ionization fraction $\chi=n_i/n_n$ where $n_n$ and $n_i$ are the neutral number density and the ion number density. The magnetic diffusion timescale $\tau_{\rm AD}$ based on $\eta_{\rm AD}$ is expected to be 

\begin{eqnarray}
\tau_{\rm AD}=\frac{L^2}{\eta_{\rm AD}}\sim5\times 10^3~{\rm yr}\biggl(\frac{L}{10^2 ~{\rm au}}\biggr)^2\biggl(\frac{B}{10^2 ~{\rm \mu G}}\biggr)^{-2}\biggl(\frac{n_n}{10^6 ~{\rm cm^{-3}}}\biggr)^{2}\biggl(\frac{\chi}{10^{-8}}\biggr)
\end{eqnarray}
where $L$ denotes a typical length. We take the disk thickness $\sim 100~\rm au$ as a typical length $L$ because the magnetic field lines are curved on that length scale. If the ionization fraction $\chi$ is smaller than $10^{-8}$, $\tau_{\rm AD}$ is smaller than acceleration timescale $\sim 10^3~\rm yr$. Therefore, magnetic acceleration may be suppressed. However, we note that there is large uncertainty in the magnitude of the ambipolar diffusion coefficient. The value of the diffusion coefficient is sensitive to the details of dust grains, which are not clearly understood.

\section{Summary}\label{sec:summary}
We investigated nonaxisymmetric late accretion onto the protoplanetary disk considering magnetic fields. We modeled the accretion as a cloudlet encounter event at a few hundred au scale. We summarize the results in the phase diagram (Figure \ref{fig:phase}) under different conditions of cloudlet size ($a_z$) and magnetic fields ($B_0$) at $t = 9008~\rm yr$. When the cloudlet is not magnetised (Gray region of Figure \ref{fig:phase}), a part of the cloudlet is ejected away from the disk after the cloudlet-disk collision and forms the sub-Keplerian spiral. When the cloudlet size is comparable or smaller than disk thickness ($a_z < 2z_{\rm s,max}$) and magnetised (Blue region of Figure \ref{fig:phase}), magnetic braking is effective and a small sub-Keplerian spiral is formed. When the cloudlet is larger than the disk thickness ($a_z > 2z_{\rm s,max}$) and magnetised (Yellow region of Figure \ref{fig:phase}), colliding cloudlet material is accelerated to a super-Keplerian velocity by magnetic fields. A part of accelerated material is gravitationally unbound. The mass of the gravitationally unbound gas is regulated by the lower limit of field strength (See subsection \ref{critical}). In earlier studies by e.g., \cite{dullemond2019} and \cite{kuffmeier2020}, spiral structures are formed by the tidal effect. In this study, we found that a sufficiently high magnetic field causes magnetic acceleration, which can rotate the spiral at super-Keplerian velocities.

\vspace{10pt}This work was supported in part by the JSPS KAKENHI grant Nos. JP18K13579, JP19K03906, JP21H04487, and JP22K14074. Numerical computations were in part carried out on Cray XC50 at Center for Computational Astrophysics, National Astronomical Observatory of Japan.

\software{ParaView \citep{paraview}, Python 3 \citep{python3} with the packages NumPy \citep{2020NumPy-Array} and Matplotlib \citep{hunter2007matplotlib}.}
%%\acknowledgments
%S.T. was supported by the JSPS KAKENHI grant Nos. JP18K13579 and JP21H04487.

\appendix
\section{Dependence of softening length of the gravitational potential}\label{appendix: softening}
\begin{figure*}
\plotone{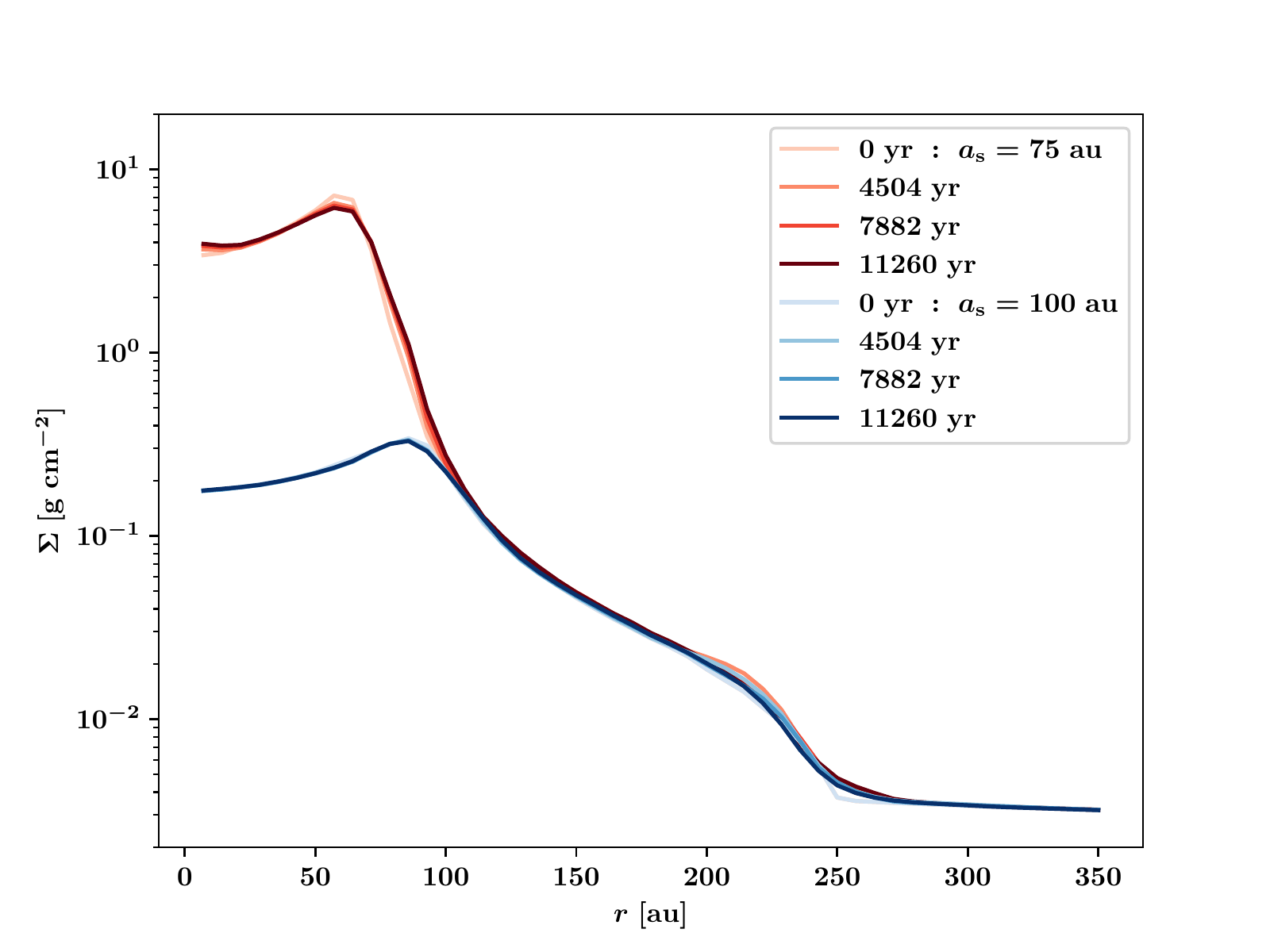}
\caption{Time evolution of the surface density distribution. Reddish lines show the results of the model with $a_s = 75~\rm au$, while bluish lines show the model with $a_s = 100~\rm au$.
}\label{fig:density_pro}
\end{figure*}

We describe the effect of softening length ($a_{\rm s}$) of the gravitational potential on the disk structure and its evolution. We perform test calculations for two models with $a_{\rm s} = 75$ and $100~\rm au$ in the absence of the cloudlet. The calculated period is 11260 yrs, which corresponds to approximately two rotational periods at $r = 250~\rm au$. Figure \ref{fig:density_pro} shows the time evolution of the surface density distribution, where one can find that both models nearly maintain the initial density distributions during the calculated periods. Therefore, the softening length does not significantly affect the disk evolution at a scale larger than the softening length.

The softening length affects the gas profile around the center. As a smaller softening length results in a deeper gravitational potential, the inner surface density is higher for the model with the smaller $a_{\rm s}$. The surface density profile outside the softening length is unchanged between the two models. We note in Figure \ref{fig:density_pro} that a peak appears in the surface density around the transition region ($r=a_{\rm s}$). This structure is an artifact produced by the artificial change in the gravitational potential and can be discerned as a ring-like structure (see panel (b) of Figure \ref{schematic view}). Nevertheless, as the cloud-disk interaction occurs in the outer part of the disk, the inner disk structure does not affect the dynamics.

\begin{figure*}
\plotone{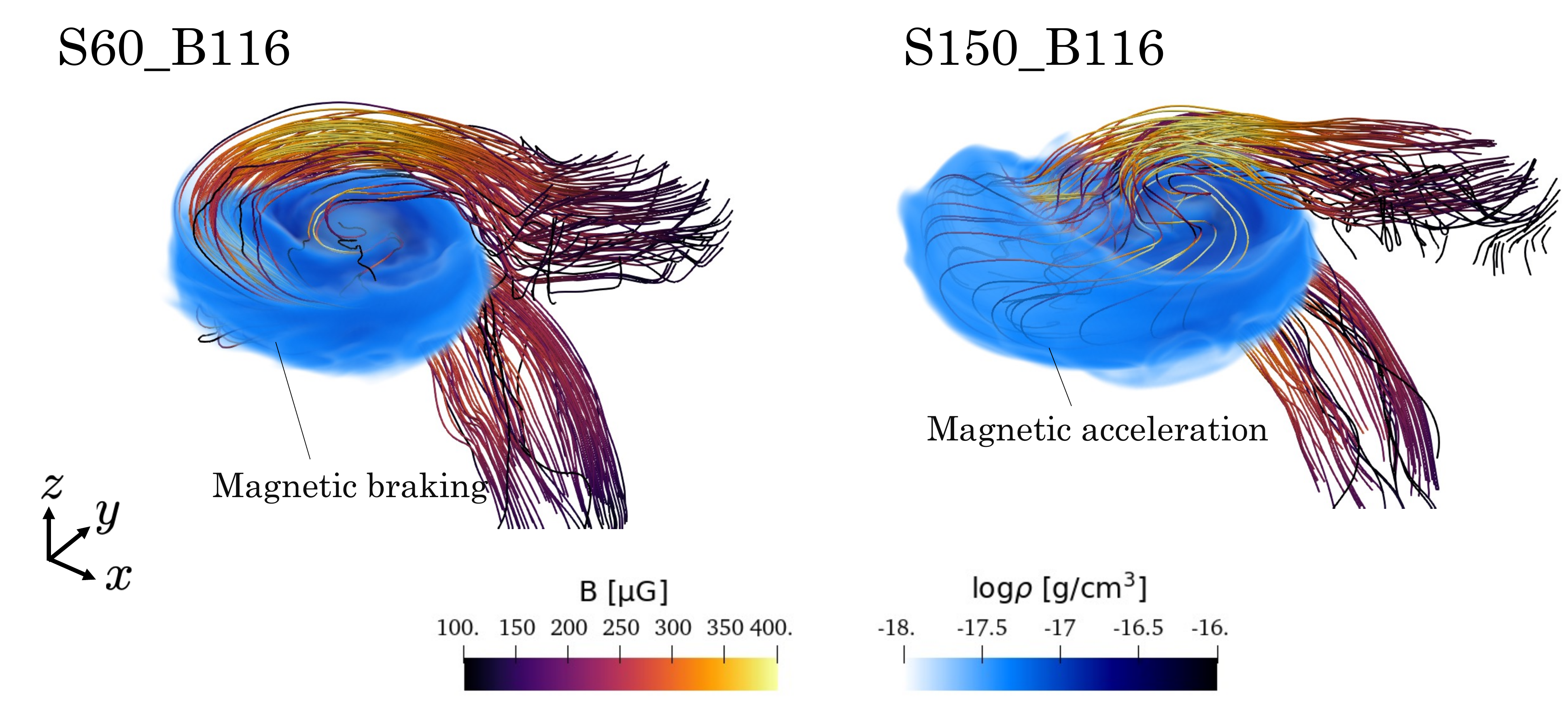}
\caption{3D structure of S60\_B116 (left) and S150\_B116 (right) with $a_{\rm s} = 75~\rm au$ at $t = 3378~\rm yr$.}\label{fig:soft_3d}
\end{figure*}

\begin{figure*}
\plotone{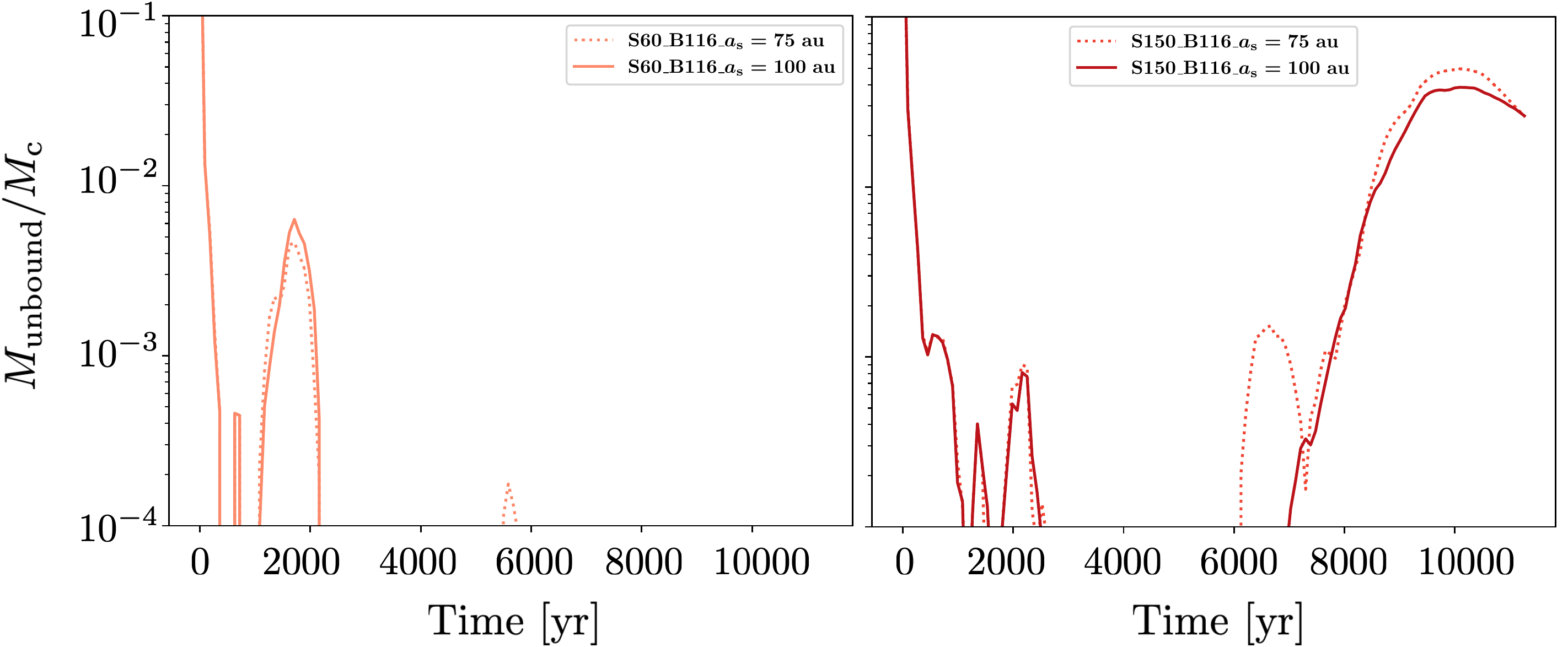}
\caption{Same as Figure \ref{fig:unbound_mass}, but for S60\_B116 and S150\_B116 with $a_{\rm s} = 75,~100~\rm au$.
}\label{fig:soft_unbound}
\end{figure*}

We also describe the effects on the cloud-disk interaction. We perform two calculations for models of S60\_B116 (magnetic braking) and S150\_B116 (magnetic acceleration) with $a_{\rm s} = 75~\rm au$, and compare these results with the models with $a_{\rm s} = 100~\rm au$. The magnetic field structure, which is essential for both magnetic braking and acceleration, is very similar between the cases with the two different softening lengths (Figure \ref{fig:soft_3d}). The mass of the gravitationally unbound gas is found to be also similar (Figure \ref{fig:soft_unbound}). Therefore, the softening length does not significantly affect the cloud-disk interaction.

\section{Impacts of boundary conditions on the cloudlet magnetic field}\label{boundary}

The models in the main text adopt fixed boundary conditions.
We describe that the top and bottom boundary conditions have little impact on the main results of this study by comparing the results with different boundary conditions. We note that the dynamics before the cloudlet collision is unaffected by the boundary conditions because the Alfv\'{e}n transit timescale outside the cloudlet (approximately a few 1,000 yrs) is longer than the infall timescale. Indeed, Figures \ref{fig:e13d} and \ref{fig:s23d} show that the magnetic field remains straight near the top and bottom boundaries approximately at 1,000 yrs when the cloudlet encounters the disk.

\begin{figure*}
\plotone{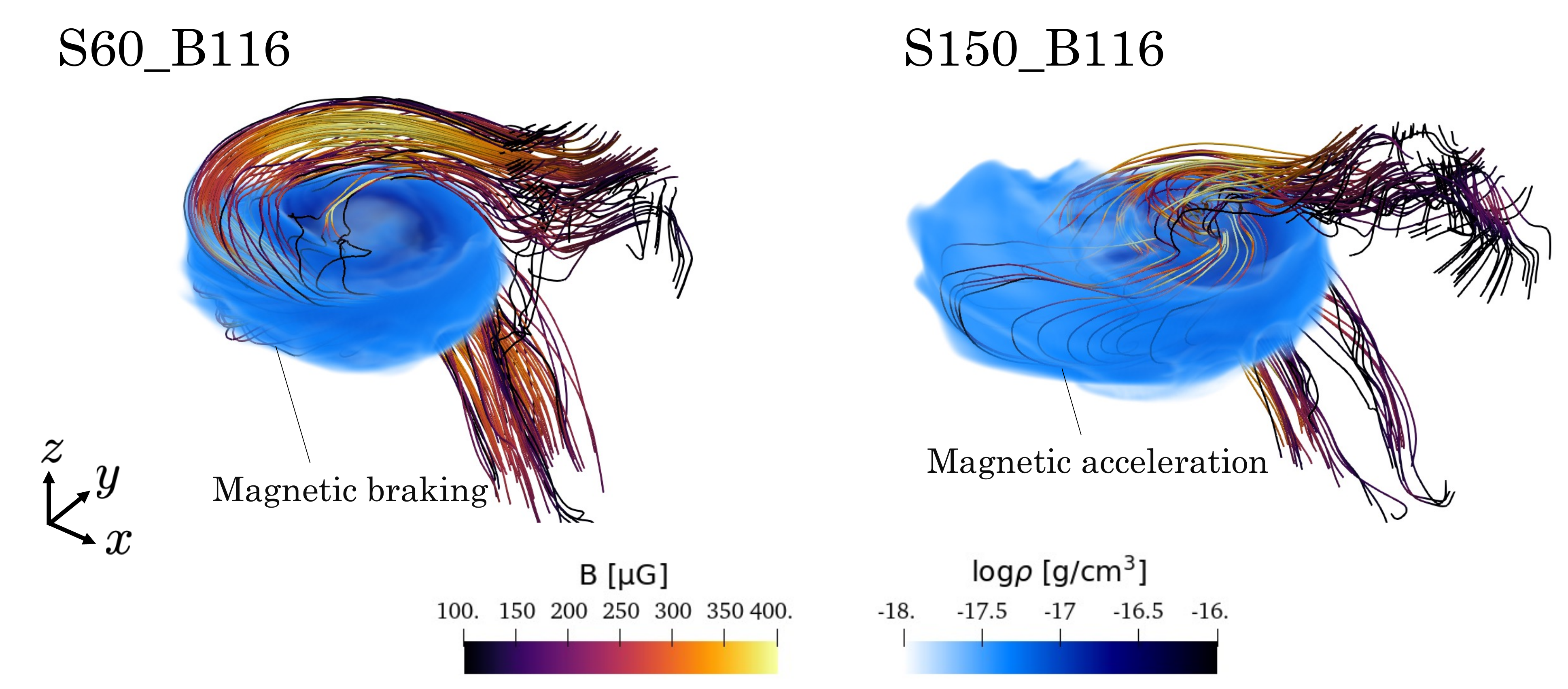}
\caption{3D structure of S60\_B116 (left) and S150\_B116 (right) with outgoing boundary conditions at $t = 3378~\rm yr$.}\label{fig:out_going_3d}
\end{figure*}

\begin{figure*}
\plotone{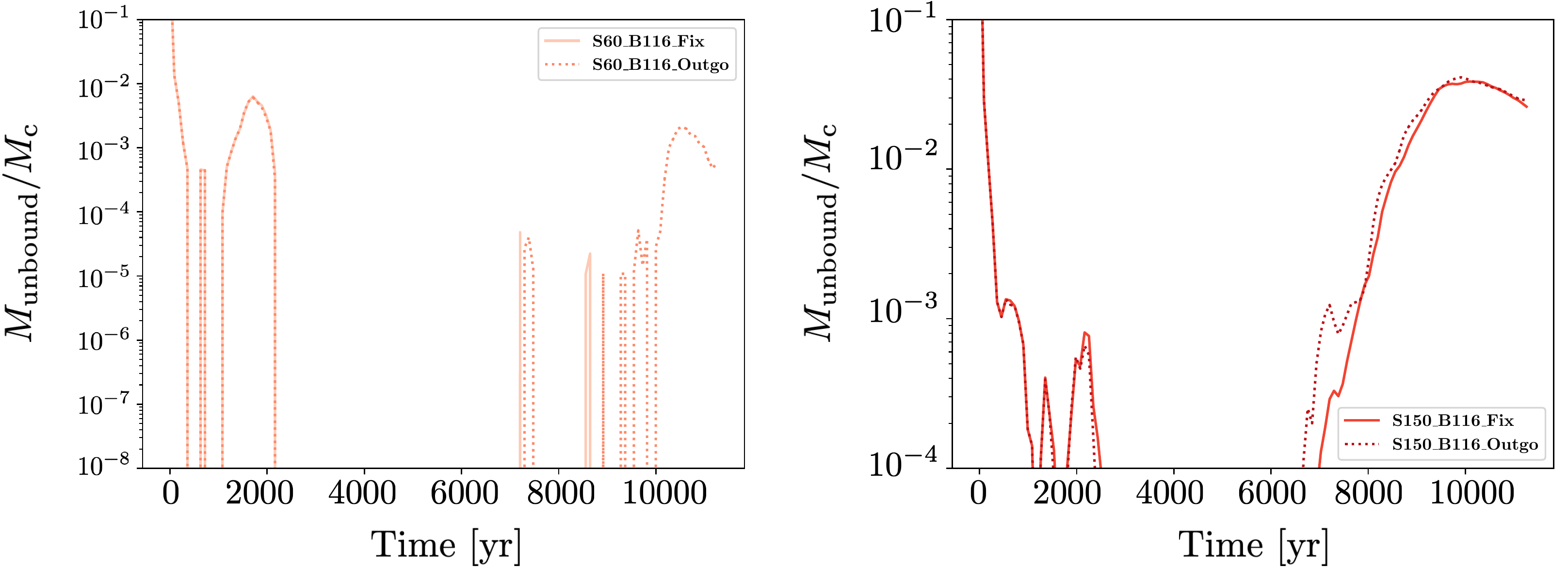}
\caption{Same as Figure \ref{fig:unbound_mass}, but for S60\_B116 and S150\_B116 with fixed boundary conditions (S60\_B116\_Fix and S150\_B116\_Fix) and outgoing boundary conditions (S60\_B116\_Outgo and S150\_B116\_Outgo). The vertical ranges of the two panels are different.}\label{fig:unbound_bnd}
\end{figure*}

We compare the results of S60\_B116 and S150\_B116 with two different boundary conditions for the top and bottom boundaries; the fixed boundary and outgoing boundary conditions. 
When the outgoing boundaries are adopted, the flows entering the simulation domain from the outside are forbidden ($v_z$ is set to zero and the zero-gradient boundary conditions are applied to the other variables). The boundaries allow the outgoing flows and apply the zero-gradient boundary conditions to all the variables of the flows.
The 3D structures for the cases of the outgoing boundary conditions are shown in Figure \ref{fig:out_going_3d}. Comparing these results with the results for the fixed boundary cases (Figures \ref{fig:e13d} and \ref{fig:s23d}), one will find that the magnetic field structures, which are essential for both magnetic braking and acceleration, are very similar to each other. We also show the mass of the gravitationally unbound gas in Figure \ref{fig:unbound_bnd}. 
For S150\_B116\_Outgo (Outgoing boundary conditions), the result is almost the same as S150\_B116\_Fix (Fixed boundary conditions). For S60\_B116\_Outgo, the mass of the gravitationally unbound gas has a small peak around 10000 yr against S60\_B116\_Fix. However, it is less than 1\% of the initial cloudlet mass, much smaller than magnetic acceleration models. Therefore, we consider that the impacts of fixed boundary conditions on the results is insignificant.

\bibliography{main}{}

\begin{thebibliography}{}
\expandafter\ifx\csname natexlab\endcsname\relax\def\natexlab#1{#1}\fi
\providecommand{\url}[1]{\href{#1}{#1}}
\providecommand{\dodoi}[1]{doi:~\href{http://doi.org/#1}{\nolinkurl{#1}}}
\providecommand{\doeprint}[1]{\href{http://ascl.net/#1}{\nolinkurl{http://ascl.net/#1}}}
\providecommand{\doarXiv}[1]{\href{https://arxiv.org/abs/#1}{\nolinkurl{https://arxiv.org/abs/#1}}}

\bibitem[{{Akiyama} {et~al.}(2019){Akiyama}, {Vorobyov}, {Liu}, {Dong}, {de
  Leon}, {Liu}, \& {Tamura}}]{akiyama2019}
{Akiyama}, E., {Vorobyov}, E.~I., {Liu}, H.~B., {et~al.} 2019, \aj, 157, 165,
  \dodoi{10.3847/1538-3881/ab0ae4}

\bibitem[{{Alcal{\'a}} {et~al.}(2017){Alcal{\'a}}, {Manara}, {Natta}, {Frasca},
  {Testi}, {Nisini}, {Stelzer}, {Williams}, {Antoniucci}, {Biazzo}, {Covino},
  {Esposito}, {Getman}, \& {Rigliaco}}]{alcala2017}
{Alcal{\'a}}, J.~M., {Manara}, C.~F., {Natta}, A., {et~al.} 2017, \aap, 600,
  A20, \dodoi{10.1051/0004-6361/201629929}

\bibitem[{{Andrews} {et~al.}(2018){Andrews}, {Terrell}, {Tripathi}, {Ansdell},
  {Williams}, \& {Wilner}}]{andrew2018}
{Andrews}, S.~M., {Terrell}, M., {Tripathi}, A., {et~al.} 2018, \apj, 865, 157,
  \dodoi{10.3847/1538-4357/aadd9f}

\bibitem[{Aota {et~al.}(2015)Aota, Inoue, \& Aikawa}]{Aota_2015}
Aota, T., Inoue, T., \& Aikawa, Y. 2015, The Astrophysical Journal, 799, 141,
  \dodoi{10.1088/0004-637x/799/2/141}

\bibitem[{{Ardila} {et~al.}(2007){Ardila}, {Golimowski}, {Krist}, {Clampin},
  {Ford}, \& {Illingworth}}]{ardila2007}
{Ardila}, D.~R., {Golimowski}, D.~A., {Krist}, J.~E., {et~al.} 2007, \apj, 665,
  512, \dodoi{10.1086/519296}

\bibitem[{Ayachit(2015)}]{paraview}
Ayachit, U. 2015, The ParaView Guide: A Parallel Visualization Application
  (Clifton Park, NY, USA: Kitware, Inc.)

\bibitem[{{Bailer-Jones} {et~al.}(2018){Bailer-Jones}, {Rybizki}, {Fouesneau},
  {Mantelet}, \& {Andrae}}]{Bailer-Jones2018}
{Bailer-Jones}, C.~A.~L., {Rybizki}, J., {Fouesneau}, M., {Mantelet}, G., \&
  {Andrae}, R. 2018, \aj, 156, 58, \dodoi{10.3847/1538-3881/aacb21}

\bibitem[{{Balbus} \& {Hawley}(1991)}]{1991ApJ...376..214B}
{Balbus}, S.~A., \& {Hawley}, J.~F. 1991, \apj, 376, 214,
  \dodoi{10.1086/170270}

\bibitem[{{Banerjee} \& {Pudritz}(2006)}]{Banerjee2006}
{Banerjee}, R., \& {Pudritz}, R.~E. 2006, \apj, 641, 949,
  \dodoi{10.1086/500496}

\bibitem[{{Beccari} {et~al.}(2010){Beccari}, {Spezzi}, {De Marchi}, {Paresce},
  {Young}, {Andersen}, {Panagia}, {Balick}, {Bond}, {Calzetti}, {Carollo},
  {Disney}, {Dopita}, {Frogel}, {Hall}, {Holtzman}, {Kimble}, {McCarthy},
  {O'Connell}, {Saha}, {Silk}, {Trauger}, {Walker}, {Whitmore}, \&
  {Windhorst}}]{beccari2010}
{Beccari}, G., {Spezzi}, L., {De Marchi}, G., {et~al.} 2010, \apj, 720, 1108,
  \dodoi{10.1088/0004-637X/720/2/1108}

\bibitem[{{Dedner} {et~al.}(2002){Dedner}, {Kemm}, {Kr{\"o}ner}, {Munz},
  {Schnitzer}, \& {Wesenberg}}]{dedner2002JCoPh}
{Dedner}, A., {Kemm}, F., {Kr{\"o}ner}, D., {et~al.} 2002, Journal of
  Computational Physics, 175, 645, \dodoi{10.1006/jcph.2001.6961}

\bibitem[{{Dullemond} {et~al.}(2019){Dullemond}, {K{\"u}ffmeier}, {Goicovic},
  {Fukagawa}, {Oehl}, \& {Kramer}}]{dullemond2019}
{Dullemond}, C.~P., {K{\"u}ffmeier}, M., {Goicovic}, F., {et~al.} 2019, \aap,
  628, A20, \dodoi{10.1051/0004-6361/201832632}

\bibitem[{{Durisen} {et~al.}(2007){Durisen}, {Boss}, {Mayer}, {Nelson},
  {Quinn}, \& {Rice}}]{Durisen2007}
{Durisen}, R.~H., {Boss}, A.~P., {Mayer}, L., {et~al.} 2007, in Protostars and
  Planets V, ed. B.~{Reipurth}, D.~{Jewitt}, \& K.~{Keil}, 607.
\newblock \doarXiv{astro-ph/0603179}

\bibitem[{{Gaia Collaboration} {et~al.}(2018){Gaia Collaboration}, {Brown},
  {Vallenari}, {Prusti}, {de Bruijne}, {Babusiaux}, {Bailer-Jones}, {Biermann},
  {Evans}, {Eyer}, {Jansen}, {Jordi}, {Klioner}, {Lammers}, {Lindegren},
  {Luri}, {Mignard}, {Panem}, {Pourbaix}, {Randich}, {Sartoretti}, {Siddiqui},
  {Soubiran}, {van Leeuwen}, {Walton}, {Arenou}, {Bastian}, {Cropper},
  {Drimmel}, {Katz}, {Lattanzi}, {Bakker}, {Cacciari}, {Casta{\~n}eda},
  {Chaoul}, {Cheek}, {De Angeli}, {Fabricius}, {Guerra}, {Holl}, {Masana},
  {Messineo}, {Mowlavi}, {Nienartowicz}, {Panuzzo}, {Portell}, {Riello},
  {Seabroke}, {Tanga}, {Th{\'e}venin}, {Gracia-Abril}, {Comoretto},
  {Garcia-Reinaldos}, {Teyssier}, {Altmann}, {Andrae}, {Audard},
  {Bellas-Velidis}, {Benson}, {Berthier}, {Blomme}, {Burgess}, {Busso},
  {Carry}, {Cellino}, {Clementini}, {Clotet}, {Creevey}, {Davidson}, {De
  Ridder}, {Delchambre}, {Dell'Oro}, {Ducourant},
  {Fern{\'a}ndez-Hern{\'a}ndez}, {Fouesneau}, {Fr{\'e}mat}, {Galluccio},
  {Garc{\'\i}a-Torres}, {Gonz{\'a}lez-N{\'u}{\~n}ez}, {Gonz{\'a}lez-Vidal},
  {Gosset}, {Guy}, {Halbwachs}, {Hambly}, {Harrison}, {Hern{\'a}ndez},
  {Hestroffer}, {Hodgkin}, {Hutton}, {Jasniewicz}, {Jean-Antoine-Piccolo},
  {Jordan}, {Korn}, {Krone-Martins}, {Lanzafame}, {Lebzelter}, {L{\"o}ffler},
  {Manteiga}, {Marrese}, {Mart{\'\i}n-Fleitas}, {Moitinho}, {Mora}, {Muinonen},
  {Osinde}, {Pancino}, {Pauwels}, {Petit}, {Recio-Blanco}, {Richards},
  {Rimoldini}, {Robin}, {Sarro}, {Siopis}, {Smith}, {Sozzetti}, {S{\"u}veges},
  {Torra}, {van Reeven}, {Abbas}, {Abreu Aramburu}, {Accart}, {Aerts},
  {Altavilla}, {{\'A}lvarez}, {Alvarez}, {Alves}, {Anderson}, {Andrei},
  {Anglada Varela}, {Antiche}, {Antoja}, {Arcay}, {Astraatmadja}, {Bach},
  {Baker}, {Balaguer-N{\'u}{\~n}ez}, {Balm}, {Barache}, {Barata}, {Barbato},
  {Barblan}, {Barklem}, {Barrado}, {Barros}, {Barstow}, {Bartholom{\'e}
  Mu{\~n}oz}, {Bassilana}, {Becciani}, {Bellazzini}, {Berihuete}, {Bertone},
  {Bianchi}, {Bienaym{\'e}}, {Blanco-Cuaresma}, {Boch}, {Boeche}, {Bombrun},
  {Borrachero}, {Bossini}, {Bouquillon}, {Bourda}, {Bragaglia}, {Bramante},
  {Breddels}, {Bressan}, {Brouillet}, {Br{\"u}semeister}, {Brugaletta},
  {Bucciarelli}, {Burlacu}, {Busonero}, {Butkevich}, {Buzzi}, {Caffau},
  {Cancelliere}, {Cannizzaro}, {Cantat-Gaudin}, {Carballo}, {Carlucci},
  {Carrasco}, {Casamiquela}, {Castellani}, {Castro-Ginard}, {Charlot},
  {Chemin}, {Chiavassa}, {Cocozza}, {Costigan}, {Cowell}, {Crifo}, {Crosta},
  {Crowley}, {Cuypers}, {Dafonte}, {Damerdji}, {Dapergolas}, {David}, {David},
  {de Laverny}, {De Luise}, {De March}, {de Martino}, {de Souza}, {de Torres},
  {Debosscher}, {del Pozo}, {Delbo}, {Delgado}, {Delgado}, {Di Matteo},
  {Diakite}, {Diener}, {Distefano}, {Dolding}, {Drazinos}, {Dur{\'a}n},
  {Edvardsson}, {Enke}, {Eriksson}, {Esquej}, {Eynard Bontemps}, {Fabre},
  {Fabrizio}, {Faigler}, {Falc{\~a}o}, {Farr{\`a}s Casas}, {Federici},
  {Fedorets}, {Fernique}, {Figueras}, {Filippi}, {Findeisen}, {Fonti},
  {Fraile}, {Fraser}, {Fr{\'e}zouls}, {Gai}, {Galleti}, {Garabato},
  {Garc{\'\i}a-Sedano}, {Garofalo}, {Garralda}, {Gavel}, {Gavras}, {Gerssen},
  {Geyer}, {Giacobbe}, {Gilmore}, {Girona}, {Giuffrida}, {Glass}, {Gomes},
  {Granvik}, {Gueguen}, {Guerrier}, {Guiraud}, {Guti{\'e}rrez-S{\'a}nchez},
  {Haigron}, {Hatzidimitriou}, {Hauser}, {Haywood}, {Heiter}, {Helmi}, {Heu},
  {Hilger}, {Hobbs}, {Hofmann}, {Holland}, {Huckle}, {Hypki}, {Icardi},
  {Jan{\ss}en}, {Jevardat de Fombelle}, {Jonker}, {Juh{\'a}sz}, {Julbe},
  {Karampelas}, {Kewley}, {Klar}, {Kochoska}, {Kohley}, {Kolenberg},
  {Kontizas}, {Kontizas}, {Koposov}, {Kordopatis}, {Kostrzewa-Rutkowska},
  {Koubsky}, {Lambert}, {Lanza}, {Lasne}, {Lavigne}, {Le Fustec}, {Le
  Poncin-Lafitte}, {Lebreton}, {Leccia}, {Leclerc}, {Lecoeur-Taibi},
  {Lenhardt}, {Leroux}, {Liao}, {Licata}, {Lindstr{\o}m}, {Lister}, {Livanou},
  {Lobel}, {L{\'o}pez}, {Managau}, {Mann}, {Mantelet}, {Marchal}, {Marchant},
  {Marconi}, {Marinoni}, {Marschalk{\'o}}, {Marshall}, {Martino}, {Marton},
  {Mary}, {Massari}, {Matijevi{\v{c}}}, {Mazeh}, {McMillan}, {Messina},
  {Michalik}, {Millar}, {Molina}, {Molinaro}, {Moln{\'a}r}, {Montegriffo},
  {Mor}, {Morbidelli}, {Morel}, {Morris}, {Mulone}, {Muraveva}, {Musella},
  {Nelemans}, {Nicastro}, {Noval}, {O'Mullane}, {Ord{\'e}novic},
  {Ord{\'o}{\~n}ez-Blanco}, {Osborne}, {Pagani}, {Pagano}, {Pailler},
  {Palacin}, {Palaversa}, {Panahi}, {Pawlak}, {Piersimoni}, {Pineau}, {Plachy},
  {Plum}, {Poggio}, {Poujoulet}, {Pr{\v{s}}a}, {Pulone}, {Racero}, {Ragaini},
  {Rambaux}, {Ramos-Lerate}, {Regibo}, {Reyl{\'e}}, {Riclet}, {Ripepi}, {Riva},
  {Rivard}, {Rixon}, {Roegiers}, {Roelens}, {Romero-G{\'o}mez}, {Rowell},
  {Royer}, {Ruiz-Dern}, {Sadowski}, {Sagrist{\`a} Sell{\'e}s}, {Sahlmann},
  {Salgado}, {Salguero}, {Sanna}, {Santana-Ros}, {Sarasso}, {Savietto},
  {Schultheis}, {Sciacca}, {Segol}, {Segovia}, {S{\'e}gransan}, {Shih},
  {Siltala}, {Silva}, {Smart}, {Smith}, {Solano}, {Solitro}, {Sordo}, {Soria
  Nieto}, {Souchay}, {Spagna}, {Spoto}, {Stampa}, {Steele},
  {Steidelm{\"u}ller}, {Stephenson}, {Stoev}, {Suess}, {Surdej}, {Szabados},
  {Szegedi-Elek}, {Tapiador}, {Taris}, {Tauran}, {Taylor}, {Teixeira},
  {Terrett}, {Teyssandier}, {Thuillot}, {Titarenko}, {Torra Clotet}, {Turon},
  {Ulla}, {Utrilla}, {Uzzi}, {Vaillant}, {Valentini}, {Valette}, {van Elteren},
  {Van Hemelryck}, {van Leeuwen}, {Vaschetto}, {Vecchiato}, {Veljanoski},
  {Viala}, {Vicente}, {Vogt}, {von Essen}, {Voss}, {Votruba}, {Voutsinas},
  {Walmsley}, {Weiler}, {Wertz}, {Wevers}, {Wyrzykowski}, {Yoldas},
  {{\v{Z}}erjal}, {Ziaeepour}, {Zorec}, {Zschocke}, {Zucker}, {Zurbach}, \&
  {Zwitter}}]{gaia2018}
{Gaia Collaboration}, {Brown}, A.~G.~A., {Vallenari}, A., {et~al.} 2018, \aap,
  616, A1, \dodoi{10.1051/0004-6361/201833051}

\bibitem[{{Gammie}(1996)}]{gammie1996}
{Gammie}, C.~F. 1996, \apj, 457, 355, \dodoi{10.1086/176735}

\bibitem[{{Garufi} {et~al.}(2021){Garufi}, {Podio}, {Codella}, {Segura-Cox},
  {Vander Donckt}, {Mercimek}, {Bacciotti}, {Fedele}, {Kasper}, {Pineda},
  {Humphreys}, \& {Testi}}]{garufi2021}
{Garufi}, A., {Podio}, L., {Codella}, C., {et~al.} 2021, arXiv e-prints,
  arXiv:2110.13820.
\newblock \doarXiv{2110.13820}

\bibitem[{{Ginski} {et~al.}(2021){Ginski}, {Facchini}, {Huang}, {Benisty},
  {Vaendel}, {Stapper}, {Dominik}, {Bae}, {M{\'e}nard}, {Muro-Arena},
  {Hogerheijde}, {McClure}, {van Holstein}, {Birnstiel}, {Boehler}, {Bohn},
  {Flock}, {Mamajek}, {Manara}, {Pinilla}, {Pinte}, \& {Ribas}}]{ginski2021}
{Ginski}, C., {Facchini}, S., {Huang}, J., {et~al.} 2021, \apjl, 908, L25,
  \dodoi{10.3847/2041-8213/abdf57}

\bibitem[{{Grady} {et~al.}(1999){Grady}, {Woodgate}, {Bruhweiler}, {Boggess},
  {Plait}, {Lindler}, {Clampin}, \& {Kalas}}]{grady1999}
{Grady}, C.~A., {Woodgate}, B., {Bruhweiler}, F.~C., {et~al.} 1999, \apjl, 523,
  L151, \dodoi{10.1086/312270}

\bibitem[{Harris {et~al.}(2020)Harris, Millman, van~der Walt, Gommers,
  Virtanen, Cournapeau, Wieser, Taylor, Berg, Smith, Kern, Picus, Hoyer, van
  Kerkwijk, Brett, Haldane, Fernández~del Río, Wiebe, Peterson,
  Gérard-Marchant, Sheppard, Reddy, Weckesser, Abbasi, Gohlke, \&
  Oliphant}]{2020NumPy-Array}
Harris, C.~R., Millman, K.~J., van~der Walt, S.~J., {et~al.} 2020, Nature, 585,
  357–362, \dodoi{10.1038/s41586-020-2649-2}

\bibitem[{{Hennebelle} \& {Ciardi}(2009)}]{hennebelle2009}
{Hennebelle}, P., \& {Ciardi}, A. 2009, \aap, 506, L29,
  \dodoi{10.1051/0004-6361/200913008}

\bibitem[{{Hennebelle} \& {Falgarone}(2012)}]{hennebelle2012}
{Hennebelle}, P., \& {Falgarone}, E. 2012, \aapr, 20, 55,
  \dodoi{10.1007/s00159-012-0055-y}

\bibitem[{{Hennebelle} \& {Fromang}(2008)}]{Hennebelle2008A&A}
{Hennebelle}, P., \& {Fromang}, S. 2008, \aap, 477, 9,
  \dodoi{10.1051/0004-6361:20078309}

\bibitem[{{Huang} {et~al.}(2020){Huang}, {Andrews}, {{\"O}berg}, {Ansdell},
  {Benisty}, {Carpenter}, {Isella}, {P{\'e}rez}, {Ricci}, {Williams}, {Wilner},
  \& {Zhu}}]{Huang2020}
{Huang}, J., {Andrews}, S.~M., {{\"O}berg}, K.~I., {et~al.} 2020, \apj, 898,
  140, \dodoi{10.3847/1538-4357/aba1e1}

\bibitem[{Hunter(2007)}]{hunter2007matplotlib}
Hunter, J.~D. 2007, Computing in science \& engineering, 9, 90

\bibitem[{{Inutsuka}(2012)}]{Inutsuka2012PTEP}
{Inutsuka}, S.-i. 2012, Progress of Theoretical and Experimental Physics, 2012,
  01A307, \dodoi{10.1093/ptep/pts024}

\bibitem[{{Joos} {et~al.}(2012){Joos}, {Hennebelle}, \& {Ciardi}}]{Joos2012A&A}
{Joos}, M., {Hennebelle}, P., \& {Ciardi}, A. 2012, \aap, 543, A128,
  \dodoi{10.1051/0004-6361/201118730}

\bibitem[{{Joos} {et~al.}(2013){Joos}, {Hennebelle}, {Ciardi}, \&
  {Fromang}}]{Joos2013A&A}
{Joos}, M., {Hennebelle}, P., {Ciardi}, A., \& {Fromang}, S. 2013, \aap, 554,
  A17, \dodoi{10.1051/0004-6361/201220649}

\bibitem[{{Kuffmeier} {et~al.}(2021){Kuffmeier}, {Dullemond}, {Reissl}, \&
  {Goicovic}}]{2021A&A...656A.161K}
{Kuffmeier}, M., {Dullemond}, C.~P., {Reissl}, S., \& {Goicovic}, F.~G. 2021,
  \aap, 656, A161, \dodoi{10.1051/0004-6361/202039614}

\bibitem[{{Kuffmeier} {et~al.}(2018){Kuffmeier}, {Frimann}, {Jensen}, \&
  {Haugb{\o}lle}}]{kuffmeier2018}
{Kuffmeier}, M., {Frimann}, S., {Jensen}, S.~S., \& {Haugb{\o}lle}, T. 2018,
  \mnras, 475, 2642, \dodoi{10.1093/mnras/sty024}

\bibitem[{{Kuffmeier} {et~al.}(2020){Kuffmeier}, {Goicovic}, \&
  {Dullemond}}]{kuffmeier2020}
{Kuffmeier}, M., {Goicovic}, F.~G., \& {Dullemond}, C.~P. 2020, \aap, 633, A3,
  \dodoi{10.1051/0004-6361/201936820}

\bibitem[{{Kuffmeier} {et~al.}(2017){Kuffmeier}, {Haugb{\o}lle}, \&
  {Nordlund}}]{kuffmeier2017}
{Kuffmeier}, M., {Haugb{\o}lle}, T., \& {Nordlund}, {\r{A}}. 2017, \apj, 846,
  7, \dodoi{10.3847/1538-4357/aa7c64}

\bibitem[{{Lam} {et~al.}(2019){Lam}, {Li}, {Chen}, {Tomida}, \&
  {Zhao}}]{Lam2019MNRAS}
{Lam}, K.~H., {Li}, Z.-Y., {Chen}, C.-Y., {Tomida}, K., \& {Zhao}, B. 2019,
  \mnras, 489, 5326, \dodoi{10.1093/mnras/stz2436}

\bibitem[{{Larson}(1969)}]{Larson1969MNRAS}
{Larson}, R.~B. 1969, \mnras, 145, 271, \dodoi{10.1093/mnras/145.3.271}

\bibitem[{{Lee} \& {Hennebelle}(2016)}]{Lee2016A&A}
{Lee}, Y.-N., \& {Hennebelle}, P. 2016, \aap, 591, A30,
  \dodoi{10.1051/0004-6361/201527981}

\bibitem[{{Li} {et~al.}(2014){Li}, {Banerjee}, {Pudritz}, {J{\o}rgensen},
  {Shang}, {Krasnopolsky}, \& {Maury}}]{Li2014prpl}
{Li}, Z.~Y., {Banerjee}, R., {Pudritz}, R.~E., {et~al.} 2014, in Protostars and
  Planets VI, ed. H.~{Beuther}, R.~S. {Klessen}, C.~P. {Dullemond}, \&
  T.~{Henning}, 173, \dodoi{10.2458/azu\_uapress\_9780816531240-ch008}

\bibitem[{{Liu} {et~al.}(2016){Liu}, {Takami}, {Kudo}, {Hashimoto}, {Dong},
  {Vorobyov}, {Pyo}, {Fukagawa}, {Tamura}, {Henning}, {Dunham}, {Karr},
  {Kusakabe}, \& {Tsuribe}}]{liu2016}
{Liu}, H.~B., {Takami}, M., {Kudo}, T., {et~al.} 2016, Science Advances, 2,
  e1500875, \dodoi{10.1126/sciadv.1500875}

\bibitem[{{Machida} {et~al.}(2007){Machida}, {Inutsuka}, \&
  {Matsumoto}}]{machida2007ApJ}
{Machida}, M.~N., {Inutsuka}, S.-i., \& {Matsumoto}, T. 2007, \apj, 670, 1198,
  \dodoi{10.1086/521779}

\bibitem[{{Machida} {et~al.}(2008){Machida}, {Inutsuka}, \&
  {Matsumoto}}]{Machida2008ApJ}
---. 2008, \apj, 676, 1088, \dodoi{10.1086/528364}

\bibitem[{{Masson} {et~al.}(2016){Masson}, {Chabrier}, {Hennebelle}, {Vaytet},
  \& {Commer{\c{c}}on}}]{2016A&A...587A..32M}
{Masson}, J., {Chabrier}, G., {Hennebelle}, P., {Vaytet}, N., \&
  {Commer{\c{c}}on}, B. 2016, \aap, 587, A32,
  \dodoi{10.1051/0004-6361/201526371}

\bibitem[{{Masunaga} \& {Inutsuka}(2000)}]{Masunaga2000ApJ}
{Masunaga}, H., \& {Inutsuka}, S.-i. 2000, \apj, 531, 350,
  \dodoi{10.1086/308439}

\bibitem[{{Matsumoto} \& {Tomisaka}(2004)}]{matsumoto2004ApJ}
{Matsumoto}, T., \& {Tomisaka}, K. 2004, \apj, 616, 266, \dodoi{10.1086/424897}

\bibitem[{{Matsumoto} {et~al.}(2019){Matsumoto}, {Asahina}, {Kudoh},
  {Kawashima}, {Matsumoto}, {Takahashi}, {Minoshima}, {Zenitani}, {Miyoshi}, \&
  {Matsumoto}}]{matsumoto2019PASJ}
{Matsumoto}, Y., {Asahina}, Y., {Kudoh}, Y., {et~al.} 2019, \pasj, 71, 83,
  \dodoi{10.1093/pasj/psz064}

\bibitem[{{Miyoshi} \& {Kusano}(2005)}]{miyoshi&kusano2005}
{Miyoshi}, T., \& {Kusano}, K. 2005, Journal of Computational Physics, 208,
  315, \dodoi{10.1016/j.jcp.2005.02.017}

\bibitem[{{Moeckel} \& {Throop}(2009)}]{moeckel2009}
{Moeckel}, N., \& {Throop}, H.~B. 2009, \apj, 707, 268,
  \dodoi{10.1088/0004-637X/707/1/268}

\bibitem[{{Mouschovias} \& {Spitzer}(1976)}]{Mouschovias1976ApJ}
{Mouschovias}, T.~C., \& {Spitzer}, L., J. 1976, \apj, 210, 326,
  \dodoi{10.1086/154835}

\bibitem[{{Nakajima} \& {Golimowski}(1995)}]{nakajima1995}
{Nakajima}, T., \& {Golimowski}, D.~A. 1995, \aj, 109, 1181,
  \dodoi{10.1086/117351}

\bibitem[{{Osterbrock}(1961)}]{1961ApJ...134..270O}
{Osterbrock}, D.~E. 1961, \apj, 134, 270, \dodoi{10.1086/147155}

\bibitem[{{Padoan} {et~al.}(2005){Padoan}, {Kritsuk}, {Norman}, \&
  {Nordlund}}]{padoan2005}
{Padoan}, P., {Kritsuk}, A., {Norman}, M.~L., \& {Nordlund}, {\r{A}}. 2005,
  \apjl, 622, L61, \dodoi{10.1086/429562}

\bibitem[{Priest(2014)}]{2014priest}
Priest, E. 2014, Magnetohydrodynamics of the Sun (Cambridge University Press),
  \dodoi{10.1017/CBO9781139020732}

\bibitem[{{Sakai} {et~al.}(2016){Sakai}, {Oya}, {L{\'o}pez-Sepulcre},
  {Watanabe}, {Sakai}, {Hirota}, {Aikawa}, {Ceccarelli}, {Lefloch}, {Caux},
  {Vastel}, {Kahane}, \& {Yamamoto}}]{sakai2016}
{Sakai}, N., {Oya}, Y., {L{\'o}pez-Sepulcre}, A., {et~al.} 2016, \apjl, 820,
  L34, \dodoi{10.3847/2041-8205/820/2/L34}

\bibitem[{{Scicluna} {et~al.}(2014){Scicluna}, {Rosotti}, {Dale}, \&
  {Testi}}]{scicluna2014}
{Scicluna}, P., {Rosotti}, G., {Dale}, J.~E., \& {Testi}, L. 2014, \aap, 566,
  L3, \dodoi{10.1051/0004-6361/201423654}

\bibitem[{{Suresh} \& {Huynh}(1997)}]{Suresh&Huynh1997}
{Suresh}, A., \& {Huynh}, H.~T. 1997, Journal of Computational Physics, 136,
  83, \dodoi{10.1006/jcph.1997.5745}

\bibitem[{{Thies} {et~al.}(2011){Thies}, {Kroupa}, {Goodwin}, {Stamatellos}, \&
  {Whitworth}}]{2011MNRAS.417.1817T}
{Thies}, I., {Kroupa}, P., {Goodwin}, S.~P., {Stamatellos}, D., \& {Whitworth},
  A.~P. 2011, \mnras, 417, 1817, \dodoi{10.1111/j.1365-2966.2011.19390.x}

\bibitem[{{Throop} \& {Bally}(2008)}]{throop2008}
{Throop}, H.~B., \& {Bally}, J. 2008, \aj, 135, 2380,
  \dodoi{10.1088/0004-6256/135/6/2380}

\bibitem[{{Tomida} {et~al.}(2013){Tomida}, {Tomisaka}, {Matsumoto}, {Hori},
  {Okuzumi}, {Machida}, \& {Saigo}}]{Tomida2013ApJ}
{Tomida}, K., {Tomisaka}, K., {Matsumoto}, T., {et~al.} 2013, \apj, 763, 6,
  \dodoi{10.1088/0004-637X/763/1/6}

\bibitem[{{Tomisaka}(2002)}]{Tomisaka2002ApJ}
{Tomisaka}, K. 2002, \apj, 575, 306, \dodoi{10.1086/341133}

\bibitem[{{Tsukamoto} {et~al.}(2015){Tsukamoto}, {Iwasaki}, {Okuzumi},
  {Machida}, \& {Inutsuka}}]{Tsukamoto2015MNRAS}
{Tsukamoto}, Y., {Iwasaki}, K., {Okuzumi}, S., {Machida}, M.~N., \& {Inutsuka},
  S. 2015, \mnras, 452, 278, \dodoi{10.1093/mnras/stv1290}

\bibitem[{Van~Rossum \& Drake(2009)}]{python3}
Van~Rossum, G., \& Drake, F.~L. 2009, Python 3 Reference Manual (Scotts Valley,
  CA: CreateSpace)

\bibitem[{Velikhov(1959)}]{Velikhov1959StabilityOA}
Velikhov, E.~P. 1959, Journal of Experimental and Theoretical Physics, 9, 995

\bibitem[{{Vorobyov}(2016)}]{Vorobyov2016}
{Vorobyov}, E.~I. 2016, \aap, 590, A115, \dodoi{10.1051/0004-6361/201628102}

\bibitem[{{Wardle}(2007)}]{2007Ap&SS.311...35W}
{Wardle}, M. 2007, \apss, 311, 35, \dodoi{10.1007/s10509-007-9575-8}

\bibitem[{{Wijnen} {et~al.}(2016){Wijnen}, {Pols}, {Pelupessy}, \& {Portegies
  Zwart}}]{wijnen2016}
{Wijnen}, T.~P.~G., {Pols}, O.~R., {Pelupessy}, F.~I., \& {Portegies Zwart}, S.
  2016, \aap, 594, A30, \dodoi{10.1051/0004-6361/201527886}

\bibitem[{{Yen} {et~al.}(2018){Yen}, {Koch}, {Manara}, {Miotello}, \&
  {Testi}}]{Yen2018}
{Yen}, H.-W., {Koch}, P.~M., {Manara}, C.~F., {Miotello}, A., \& {Testi}, L.
  2018, \aap, 616, A100, \dodoi{10.1051/0004-6361/201732196}

\end{thebibliography}
\bibliographystyle{aasjournal}

\end{document}